\documentclass[lettersize,journal]{IEEEtran}
\usepackage{amsmath,amsfonts}
\usepackage{algorithm}
\usepackage{algpseudocode}

\usepackage{array}
\usepackage[caption=false,font=normalsize,labelfont=sf,textfont=sf]{subfig}
\usepackage{textcomp}
\usepackage{stfloats}
\usepackage{url}
\usepackage{verbatim}
\usepackage{graphicx}
\usepackage{cite}
\usepackage{tabularx}
\usepackage{booktabs}
\usepackage{comment}
\usepackage{xcolor}

\begin{document}
	
	\title{An Adaptive Environment-Aware Transformer Autoencoder for UAV-FSO with Dynamic Complexity Control}
	
	\author{Han Zeng$^{\ast}$, Haibo Wang$^{\ast}$,\IEEEmembership{ Member,~IEEE,} Kan Wang, Xutao Yu,  Zaichen Zhang,~\IEEEmembership{Senior Member,~IEEE}
		\thanks{$^{\ast}$Han Zeng and Haibo Wang contributed equally to this work.}%
		\thanks{This work is supported by NSFC projects (62471126, 623B2017 and 61960206005),
			the Fundamental Research Funds for the Central Universities 2242022k60001, Jiangsu Key R\&D Program Project BE2023011-2, the research fund of National Mobile Communications Research Lab. 2025A03. (Corresponding author: Zaichen Zhang.)}
		\thanks{
			Han Zeng, Haibo Wang and Zaichen Zhang are with the National Mobile Communications Research Laboratory, Southeast University, Nanjing 210096, P.R.China, and they are also with the Purple Mountain Laboratory, Nanjing 211111, P. R. China. (e-mail: 230228224@seu.edu.cn, haibowang@seu.edu.cn, zczhang@seu.edu.cn)
			
			Kan Wang is with the Purple Mountain Laboratory, Nanjing 211111, P. R. China. (e-mail:  wangkan@pmlabs.com.cn)
			
			Xutao Yu is with the State Key Lab of Millimeter Waves, Southeast University, Nanjing 210096, P. R. China, and also with the Purple Mountain Laboratory, Nanjing 211111, P. R. China.  (e-mail: yuxutao@seu.edu.cn)}}
	
	\markboth{Journal of \LaTeX\ Class Files,~Vol.~14, No.~8, August~2021}%
	{Shell \MakeLowercase{\textit{et al.}}: A Sample Article Using IEEEtran.cls for IEEE Journals}

	\maketitle
	
	\begin{abstract}
		
		The rise of sixth-generation (6G) wireless networks imposes stringent requirements on UAV-assisted Free Space Optical (FSO) communications, where atmospheric turbulence and UAV-induced vibrations lead to highly dynamic and complex channel conditions. While end-to-end autoencoder-based communication frameworks enable data-driven transceiver adaptation to such impairments, existing designs remain largely generic and lack the environment awareness and computational flexibility required for practical UAV-FSO scenarios. To address this, we propose AEAT-AE (Adaptive Environment-aware Transformer Autoencoder), a Transformer-based framework that integrates environmental parameters into both encoder and decoder via a cross-attention mechanism. Moreover, AEAT-AE incorporates a Deep Q-Network (DQN) that dynamically selects which layers of the Transformer autoencoder to activate based on real-time environmental inputs, balancing performance and computational cost. Simulation results demonstrate that AEAT-AE outperforms conventional methods in bit error rate while achieving a favorable performance–complexity trade-off under varying channel conditions, representing a tailored solution for next-generation UAV-FSO communications.

	\end{abstract}
	
	
	\section{Introduction}
	
	\IEEEPARstart{A}{s} the development of sixth-generation (6G) wireless networks advances, requirements for ultra-high data rates, massive connectivity, and low latency continue to intensify. These demands, driven by emerging applications such as immersive multimedia services, distributed edge computing, and industrial Internet of Things (IoT) systems, further exacerbate the limitations of the already congested radio frequency (RF) spectrum \cite{6G1}. In this context, Free Space Optical (FSO) communication, as a key technology within Optical Wireless Communication (OWC), has emerged as a promising solution for high-capacity wireless links by virtue of its extremely high bandwidth, immunity to RF interference, low latency, and enhanced security \cite{ref1,ref2,ref3}.
	
	Building upon these advantages, the integration of Unmanned Aerial Vehicles (UAVs) with FSO communication further extends the applicability of OWC in 6G networks by enabling flexible and rapidly deployable communication architectures. In particular, UAV-assisted FSO systems are well suited for scenarios requiring on-demand coverage extension, such as disaster recovery operations with damaged ground infrastructure, remote area communications lacking fiber backhaul, and dense urban environments with high-capacity backhaul demands \cite{ref4p,ref5p,ref6}. By mounting FSO terminals on UAV platforms, reconfigurable line-of-sight links can be established to support backhaul, relaying, and direct access services \cite{ref7,ref8,ref9,ref10,ref11}.
	
	UAV-FSO systems offer substantial advantages in terms of high-capacity, secure, and flexible communication. However, they are also affected by tightly coupled impairments that complicate reliable link maintenance. For example, platform jitter and UAV motion can cause persistent beam misalignment~\cite{ref12}, while atmospheric turbulence and weather-induced fading further degrade the optical signal quality~\cite{ref13}. Although analytical channel models provide useful insights into these effects, many system-level optimization problems—especially those requiring end-to-end or coordinated adaptation—remain challenging to solve without strong simplifying assumptions. These practical difficulties have motivated the exploration of learning-based techniques that can capture interactions across multiple components of the communication chain.
	
\begin{figure*}[!t]
		\centering
		\includegraphics[width=7in]{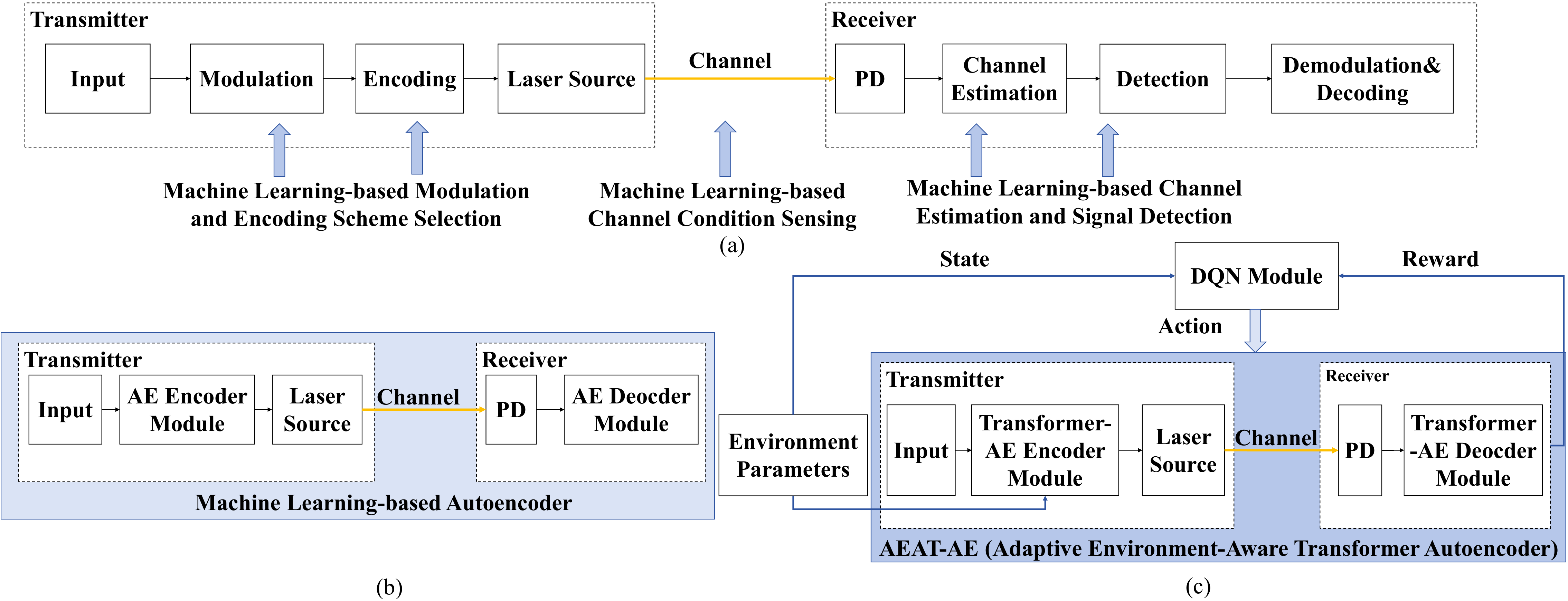}
		\caption{Comparison of schemes for applying machine learning to the transmission process of communication systems: (a) machine learning optimization for specific modules; (b) The system model of traditional end-to-end autoencoder; (c) The system model proposed in this work.}
		\label{Compare}
	\end{figure*}

	Most existing data-driven studies, however, focus on individual blocks of the communication pipeline, as illustrated in Fig.~\ref{Compare}(a). For instance, support vector machines (SVMs) and deep neural networks have been applied to enhance detector robustness under varying channel conditions~\cite{ref14, ref14p0, ref15}, while convolutional neural networks (CNNs) have been used to predict channel fluctuations for proactive adjustment~\cite{ref16, ref16p}. These approaches, by treating each module in isolation, often struggle to coordinate transmitter–receiver adaptation in rapidly changing UAV-FSO environments.
	
	To overcome these limitations, recent research has adopted end-to-end learning strategies. Autoencoder (AE) architectures jointly model the encoder, channel, and decoder within a single differentiable framework, as illustrated in Fig.~\ref{Compare}(b). By optimizing all components simultaneously, AE-based transceivers can adapt to complex channel behaviors that are difficult to describe analytically, enabling effective system-level optimization under dynamic conditions.

	\subsection{Prior Works}
	
	\begin{table*}
		\centering
		\caption{Related Works}
		\begin{tabularx}{\linewidth}{|>{\centering\arraybackslash}m{1.5cm}|>{\centering\arraybackslash}m{1.2cm}|>{\centering\arraybackslash}m{3cm}|>{\centering\arraybackslash}m{5.5cm}|>{\centering\arraybackslash}m{2cm}|>{\centering\arraybackslash}X|}
			\hline
			\textbf{Ref.} & \textbf{RF or Optical} & \textbf{Main Propose} & \textbf{Main Contribution} & \textbf{Main Network or Model} & \textbf{Main Comparative Indicators}\\
			\hline
			\cite{Related1} (2017) & RF & End-to-End Optimization & Introduces the use of AE for end-to-end communication systems & DNN (FCN) & BLER\\
			\hline
			\cite{Related8} (2018) & RF & Model-Free Channel AE &Optimizes AE without channel gradients & DNN (FCN) & BER\\
			\hline
			\cite{Related9} (2018) & Optical (OWC) & End-to-End Optimization & Introduces autoencoder-based optical wireless communication systems
			& DNN (FCN) & BLER\\
			\hline
			\cite{Related3} (2019) & RF & Model-Free Channel AE & Introduces a RL-based AE without gradient backpropagation
			& RL & BLER, PSNR\\
			\hline
			\cite{Related11} (2019) & Optical (OWC) & End-to-End Optimization & Introduces autoencoder-based transceivers
			& DNN (FCN) & BLER\\
			\hline
			\cite{Related2} (2020) & RF & Model-Free Channel AE & Introduces an end-to-end communication system using conditional GANs to model wireless channels & CNN \& GAN & BLER \& BER\\
			\hline
			\cite{Related7} (2020) & RF & JSCC with Feedback & Introduces CNN-based AE using multiple encoder-decoder pairs for progressive transmission & CNN & PSNR\\
			\hline
			\cite{Related10} (2020) & Optical (OWC) & End-to-End Optimization & Introduces AE based asymmetrically-clipped optical OFDM communication system & ReLU+IFFT/FFT & BER\\
			\hline
			\cite{Related13} (2022) & Optical (FSO) & End-to-End Optimization & Introduces AE-based free space optical communication systems & DNN (FCN) & BER\\
			\hline
			\cite{Related12} (2023) & Optical (FSO) & End-to-End Optimization & Introduces learning-based autoencoder for multiple access and interference channels & DNN (FCN) & BER\\
			\hline
			\cite{Related4} (2024) & Optical (FSO) & End-to-End Optimization & Introduces an end-to-end AE in non-differentiable Poisson channel & DNN (FCN) & BER\\
			\hline
			\cite{Related5} (2024) & Optical (OWC) & End-to-End Optimization & Introduces a differential AE for OWC systems & DNN (FCN) & BER\\
			\hline
		\end{tabularx}
		\label{Related works}
	\end{table*}

	Building on this end-to-end paradigm, existing AE-based communication systems have evolved along three main research directions, as summarized in Table~\ref{Related works}. Early works (2017–2019) proposed basic AE architectures using fully connected networks (FCNs) in both RF~\cite{Related1,Related8} and optical~\cite{Related9,Related11} domains. These studies focused on performance metrics such as bit error rate (BER) and block error rate (BLER). Later developments (2019–2020) explored new training paradigms, including reinforcement learning~\cite{Related3} and GAN-based channel modeling~\cite{Related2}. They also introduced architectural advances such as CNN-based joint source-channel coding (JSCC)~\cite{Related7} and optical OFDM~\cite{Related10}. More recent efforts (2020–2024) have targeted FSO-specific challenges, including Poisson channel non-differentiability~\cite{Related4}, multi-user interference~\cite{Related12}, turbulence resilience~\cite{Related13}, and differential coding~\cite{Related5}.

	While AE-based frameworks have significantly advanced end-to-end optimization, directly applying them to UAV-FSO systems remains challenging. The dynamic behavior of UAV-FSO links differs substantially from conventional RF or ground FSO channels, which may limit the effectiveness of standard approaches. Most existing studies use standard channel models and simulated datasets for reproducibility, but they face limitations when applied to UAV-FSO scenarios. First, they are typically developed for generic RF or FSO channels rather than UAV-FSO links. Second, their black-box architectures offer limited adaptability to environmental variations. Third, environmental parameters are often included superficially rather than embedded as core components of the encoding process. These limitations are amplified in UAV-FSO links, where turbulence, mobility, and misalignment jointly produce highly dynamic and nonlinear channel behavior, making effective adaptation essential.

	\subsection{Our Contributions}
	
	To address these limitations in UAV-FSO settings, we propose the Adaptive Environment-Aware Transformer Autoencoder (AEAT-AE), a framework specifically tailored for such systems. As shown in Fig. 1 (c), AEAT-AE performs end-to-end optimization by jointly modeling the transmitter and receiver, while embedding environmental parameters directly into both encoding and decoding. Its symmetric encoder-decoder structure enables the encoder to extract compact latent representations and the decoder to reconstruct signals under severe channel distortions. Multi-head attention captures diverse channel characteristics, supporting robust adaptation to turbulence, UAV mobility, and beam misalignment. However, this powerful architecture comes with high computational cost, particularly under highly dynamic channel conditions.
	
	To mitigate this computational burden, we further enhance AEAT-AE's adaptability by integrating a Deep Q-Network (DQN)-based dynamic layer selection mechanism. The DQN controller learns, during inference, which Transformer layers are necessary under the current environmental state, enabling the model to scale its computational load according to link conditions~\cite{DQL1,DQL2,QL3}. This approach provides a principled alternative to heuristic layer selection and is well suited to the highly variable nature of UAV-FSO channels.
	
	The main contributions of this work are summarized below:
	
	(1) Environment-Integrated Transformer AE Design: We propose a Transformer-based autoencoder architecture, AEAT-AE, which explicitly integrates environmental parameters into both encoding and decoding. This design enhances the model's adaptability to diverse UAV-FSO channel conditions, addressing the limitations of traditional AE models that treat environmental information as isolated inputs.
	
	(2) Dynamic Inference via DQN-based Layer Selection: We introduce a DQN-based dynamic layer selection strategy that adaptively determines which Transformer layers are activated during inference, based on environmental conditions and system constraints. This approach enables adaptive control of model complexity without retraining, reducing computational cost under stable conditions while maintaining adaptability in dynamic environments.
	
	(3) Improved Performance–Complexity Trade-off: Through extensive experiments and ablation studies, we demonstrate that AEAT-AE improves channel robustness and BER compared with FCN- and CNN-based baselines. The DQN-based dynamic layer selection adapts computation to environmental conditions, achieving a favorable performance–complexity trade-off without degrading accuracy.

	\subsection{Organization}
	
	The rest of the paper is organized as follows. Section II introduces the system and channel models for UAV-FSO communications. Section III presents the proposed AEAT-AE framework, focusing on the Transformer-based encoder-decoder architecture and its end-to-end design principles. Section IV describes the DQN-based dynamic layer selection mechanism and the overall training and deployment procedure. Section V provides experimental results and performance evaluation. Section VI discusses the results and outlines future research directions. Finally, Section VII concludes the paper.
	
	\section{System Model} 
	
	\subsection{AE-Based UAV-FSO System Model}
	
	\begin{figure}[!t]
		\centering
		\includegraphics[width=3.5in]{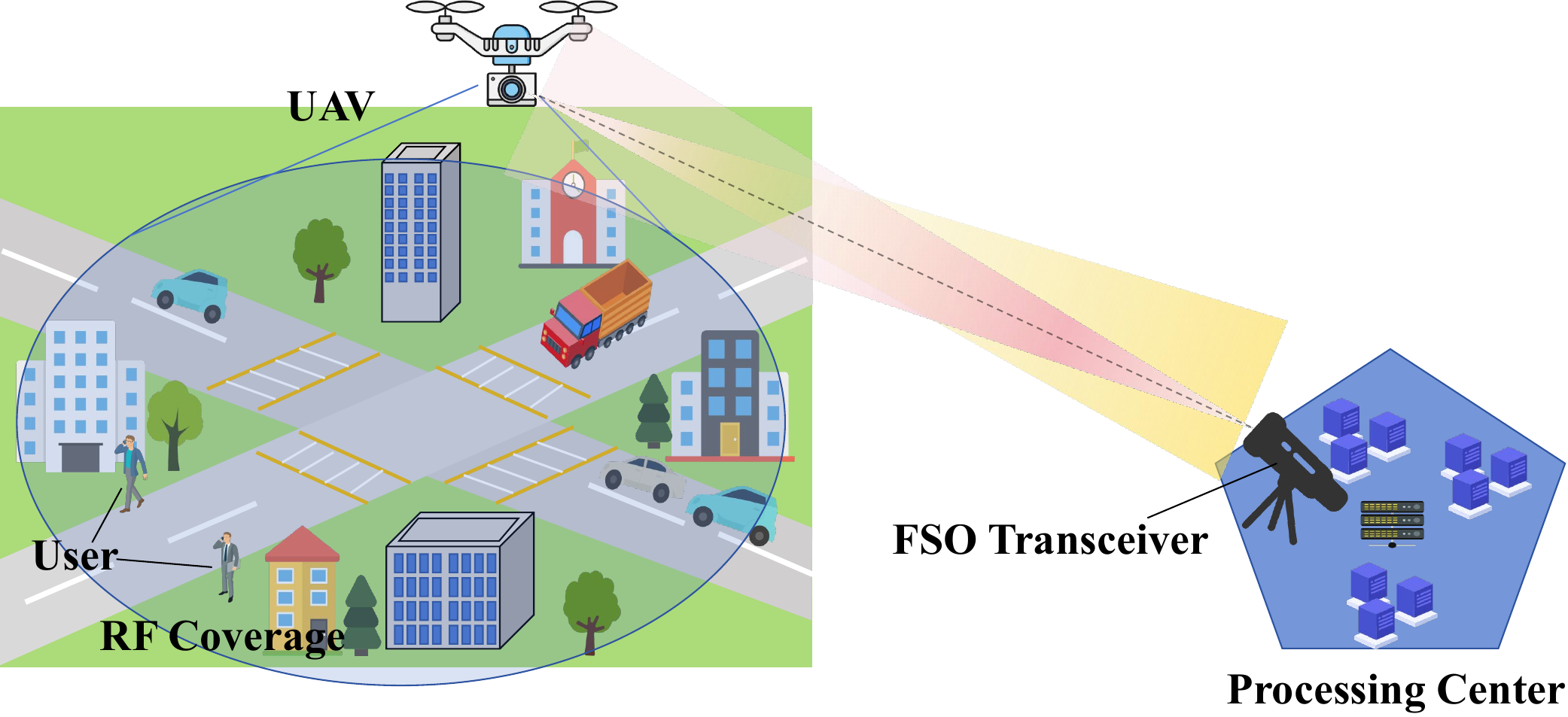}
		\caption{One representative application scenario of UAV-FSO systems.}
		\label{Background}
	\end{figure}
	
	Fig.~\ref{Background} illustrates a typical UAV-FSO communication scenario, where a ground data center establishes a high-capacity optical wireless link to a UAV. The UAV serves as a mobile relay, while providing coverage to users. The optical channel is subject to rapid fluctuations caused by UAV-induced jitter, atmospheric turbulence, and pointing errors. These challenges motivate the development of adaptive, data-driven transceiver designs that can maintain reliable high-capacity FSO links.
	
	Given an input bitstream $\mathbf{s}\in\{0,1\}^N$, where $N$ denotes the bitstream length, the proposed autoencoder (AE) jointly learns the transmitter and receiver mappings in an end-to-end manner. The encoder maps $\mathbf{s}$ to a latent vector $\mathbf{x}\in\mathbb{R}^{K\times 1}$, where $K$ denotes the length of the continuous signal vector. To comply with the physical constraints of intensity modulation and direct detection (IM/DD), a sigmoid activation is applied at the encoder output to ensure non-negativity of each element of $\mathbf{x}$. Furthermore, normalization is employed to satisfy the average optical power constraint $\mathbb{E}[\|\mathbf{x}\|_1/K] \leq P_{\text{avg}}$, while the peak optical intensity is inherently bounded by the activation function, ensuring compliance with practical transmitter limitations and eye-safety requirements.
	
	The UAV-FSO channel is modeled as a standard single-beam IM/DD link. The received signal $\mathbf{y} \in \mathbb{R}^{K \times 1}$ is given by
	\begin{equation}
		\mathbf{y} = \eta_e P_t\, h\, \mathbf{x} + \mathbf{w},
	\end{equation}
	
	\noindent where $\eta_e$ denotes the photodetector responsivity, $P_t$ is the transmitted optical power, and $h\in\mathbb{R}$ is the composite channel gain capturing atmospheric attenuation, turbulence-induced fading, pointing errors, and UAV misalignment effects. The additive noise vector $\mathbf{w}\sim\mathcal{N}(0,\sigma_w^2\mathbf{I}_K)$ models the combined impact of shot noise and thermal noise at the IM/DD receiver. Following common practice in FSO literature\cite{ref5p,Ref5pp,Ref5ppp}, the noise is modeled as signal-independent Gaussian, which is accurate when background and thermal noise dominate.

	At the receiver, the AE decoder learns to map the received signal back to the original bitstream $\hat{\mathbf{s}}\in[0,1]^N$, representing the posterior probabilities of the transmitted bits. The system is optimized using the binary cross-entropy (BCE) loss, which directly aligns the end-to-end training objective with the bit-level transmission task.
	
	\begin{equation}
		\min \sum_{n=1}^{N} \big(s_n\log(\hat{s}_n) + (1-s_n)\log(1-\hat{s}_n)\big).
	\end{equation}
	
	In practical UAV-FSO systems, on--off keying (OOK) is widely adopted due to its simplicity, robustness, and compatibility with IM/DD optical transmitters under strict size, weight, and power constraints. In the proposed AE transceiver, the encoder output $\mathbf{x}$ is first processed to satisfy physical-layer constraints and then discretized into an OOK-modulated optical waveform. This implementation choice ensures consistency between the learned AE-based transceiver and realistic optical wireless hardware.

	\subsection{Channel Fading Model for UAV-FSO Systems}
	
	UAV-FSO channels are highly sensitive to beam misalignment, UAV-induced motion, and atmospheric turbulence, leading to rapid fluctuations in the channel coefficient $h$. Accurately capturing these dynamics requires a model that balances realism with analytical tractability. In this work, we adopt a widely used fading model that combines internal disturbances (e.g., platform vibration, mechanical jitter) and external factors (e.g., wind, turbulence) into a single stochastic jitter parameter. This approach has been extensively validated in the UAV-FSO literature~\cite{ref3,ref4p,ref5p,ref7,Add_1,Add_2,ref7pp,ref7ppp,ref13} and enables controlled, repeatable performance evaluation.
	
	We focus on evaluating the adaptability of the Transformer-AE framework under dynamic conditions using a simulation-based approach for its analytical clarity. Instead of proposing a brand new channel model, we leverage an established abstraction to investigate Transformer-based adaptation in UAV-FSO systems. To capture the composite effects involved, the overall channel coefficient is modeled as:
	
	\begin{equation}
		h=h_{l} h_{a} h_{p} h_{\mathrm{AoA}},
	\end{equation}
	\noindent where $h_{l}$ is the atmospheric attenuation resulting from the scattering and absorption phenomena, modeled by the Beer-Lambert law: $h_l=\exp \left(-Z \xi \right)$, with $Z$ as the link distance and $\xi$ as the scattering coefficient influenced by visibility distance $V_d$. $h_{a}$ is the atmospheric turbulence fading, $h_{p}$ is the pointing error fading and $h_{\mathrm{AoA}}$ denotes the link interruption due to radial angle of arrival (AoA) fluctuations.

	\subsubsection{Atmospheric Turbulence $h_a$}
	
	This fading is caused by temperature and density fluctuations in the atmosphere. We choose the Málaga distribution to model both weak and strong turbulence conditions~\cite{ref7p}. Here, $h_a$ denotes the turbulence-induced fading coefficient. The PDF of $h_a$ is:
	
	\begin{equation}
		f_{h_a}(h_a) = A \sum_{m=1}^{\beta} \left( a_m \, h_a^{\frac{\alpha+m}{2}-1} \right)
			K_{\alpha-m}\!\left( 2\sqrt{\frac{\alpha\beta h_a}{b\beta+\Omega'}} \right), \quad h_a \ge 0,
		\label{eq14}
	\end{equation}
	
	\begin{equation}
		\left\{
		\begin{array}{l}
			A = \frac{2\alpha^{\frac{\alpha}{2}}}{b^{\,1+\frac{\alpha}{2}} \Gamma(\alpha)}
			\left(\frac{b\beta}{b\beta+\Omega'}\right)^{\beta+\frac{\alpha}{2}}, \\
			a_m = \binom{\beta-1}{m-1}
			\frac{(b\beta+\Omega')^{1-\frac{m}{2}}}{(m-1)!}
			\left(\frac{\Omega'}{b}\right)^{m-1}
			\left(\frac{\alpha}{\beta}\right)^{\frac{m}{2}}
		\end{array}
		\right. ,
		\label{eq15}
	\end{equation}
	
	\begin{equation}
		\Omega' = \Omega + \rho\,2b_0 + 2\sqrt{2b_0\Omega\rho}\cos(\phi_A-\phi_B),
		\label{eq17}
	\end{equation}
	
	\begin{equation}
		\rho = (0.55\, C_n^2 \lambda^2 Z)^{-3/5},\\
		b = 2b_0 (1-\rho)
		\label{eq24}
	\end{equation}
	
	\noindent where  
		$\alpha$ is the number of large-scale turbulence cells,  
		$\beta$ is the fading parameter,  
		$K_v(\cdot)$ is the modified Bessel function of the second kind of order $v$,  
		$\phi_A-\phi_B$ is the phase difference (rad),  
		$\rho$ is the amount of scattering power coupled to the LoS component,  
		$2b_0$ is the average scattering power (W),  
		$\lambda$ is the beam wavelength (m),  
		$C_n^2$ is the refractive index structure parameter,  
		$Z$ is the propagation distance (m),  
		$\Omega$ is the average power of LoS component (W).

	\subsubsection{Point Error due to Transmitter and UAV Jitter $h_p$}
	
	UAV mobility and mechanical vibrations of platforms cause pointing errors, leading to misalignment between the transmitter and receiver. For a Gaussian beam at the receiver, power loss due to radial displacement $r_d$ is: 
	
	\begin{equation}
		h_{p} \approx A_{0} \exp \left(-\frac{2 r_{d}^{2}}{w_{z_{eq}}^{2}}\right)
	\end{equation}
	
	\noindent where $A_0$ is the power fraction at zero pointing error and can be presented as:
	
	\begin{equation}
		A_{0}=\operatorname{erf}(v)^{2}, \quad v=\sqrt{\frac{\pi}{2}} \frac{r_{a}}{w_{z}},
	\end{equation}
	
	\noindent where $\text{erf}(\cdot)$ represents the error function, $w_{z_{eq}}$ denotes the equivalent beamwidth,
	$w_z$ is the Gaussian beam waist at propagation distance $Z$ (computed from $w_{oz}$ via (\ref{eq22})):
	
	\begin{equation}
		w_{z_{eq}}=\frac{w_z^2 \sqrt{2} erf(v)}{2v \exp(-v^2)}
	\end{equation}
	
	\begin{equation}
		w_z \approx w_{oz} \sqrt{1+\Theta \left(\frac{\lambda Z}{\pi w^2_{oz}}\right)^2}, \quad \Theta=1+\frac{2w^2_{oz}}{{\rho}^2},
		\label{eq22}
	\end{equation}
	
	\noindent
	$w_{oz}$ is the beam width at $Z=0$ (m), 
	$r_a$ is the receiver lens radius (m).  
	
	UAV jitter, caused by factors like hardware structure, payload, and gimbal control, increases radial displacement from the beam center to the receiving aperture. This jitter is often modeled as a Gaussian-distributed random variable, and the total radial displacement $r_d=\sqrt{x_d^2+y_d^2}$ follows a Beckmann distribution\cite{ref7p,ref7pp,ref7ppp}. Since its PDF lacks a closed-form expression, a modified Rayleigh distribution is used for approximation:
	
	\begin{equation}
		f_{h_p}(h_p)=\frac{g^2}{A^{g^2}_0} h^{g^2-1}_p, 0\leq h_p \leq A_0
		\label{eq25}
	\end{equation}
	
	\noindent where 
	$g=\frac{w_{z_{eq}}}{2\sigma_s}$ is the pointing error coefficient, $\sigma_s$ is the pointing error standard deviation.
	
	\subsubsection{Fading due to AoA Fluctuations caused by UAV Motion $h_{\mathrm{AoA}}$}
	
	In UAV-to-Ground (U2G) FSO scenarios, the deviation in angle of arrival (AoA) is defined as $\theta_{\mathrm{AoA}} = \sqrt{\theta_{tx}^2 + \theta_{ty}^2}$, where 
	$\theta_{tx}$ and $\theta_{ty}$ denote the boresight bias angles caused by UAV jitter, turbulence, and platform disturbances (radian). A commonly adopted model in UAV-FSO studies~\cite{ref7p,ref7pp,ref7ppp} assumes that if the AoA exceeds the receiver’s field of view (FoV), the signal is lost and $h_{\mathrm{AoA}} = 0$; otherwise, $h_{\mathrm{AoA}} = 1$:
	
	\begin{equation}
		h_{\mathrm{AoA}}=\left\{
		\begin{array}{ll}
			1, & \theta_{\mathrm{AoA}} \leq \theta_{FoV} \\[1mm]
			0, & \theta_{\mathrm{AoA}} > \theta_{FoV}
		\end{array}
		\right.
	\end{equation}
	
	To model its stochastic behavior, we adopt a Beckmann-based distribution:
	
	\begin{equation}
		f_{h_{\mathrm{AoA}}}(h_{\mathrm{AoA}}) = S \delta(h_{\mathrm{AoA}}) + (1 - S) \delta(h_{\mathrm{AoA}} - 1),
	\end{equation}
	
	\noindent where $S = \exp\left(-\frac{\theta^2_{FoV}}{2\sigma_a^2}\right)$, 
	$\sigma_a$ represents the standard deviation of angular jitter (radian).
	
	To explicitly characterize the stochastic nature of AoA fluctuations, the distribution of $\theta_{\mathrm{AoA}}$ is approximated as:
	
	\begin{equation}
		f_{\theta_{\mathrm{AoA}}}\left(\theta_{\mathrm{AoA}}\right) \approx \frac{\theta_{\mathrm{AoA}}}{\sigma_{\mathrm{AoA}}^{2}} \exp \left(-\frac{\theta_{\mathrm{AoA}}^{2}}{2 \sigma_{\mathrm{AoA}}^{2}}\right), \quad \theta_{\mathrm{AoA}} \geq 0,
	\end{equation}
	
	\noindent where $\sigma_a^2$ encapsulates the UAV’s orientation instability. For U2G links, it is given by:
	
	\begin{equation}
		\sigma_{a}^{2} = \left(\frac{3 \theta_{tx}^{\prime 2} \sigma_{txo}^{4} + 3 \theta_{ty}^{\prime 2} \sigma_{tyo}^{4} + \sigma_{txo}^{6} + \sigma_{tyo}^{6}}{2}\right)^{\frac{1}{3}},
	\end{equation}
	
	\noindent where 
	$\sigma_{txo}$ and $\sigma_{tyo}$ denote the standard deviations of the transmitter UAV’s orientation in the $x$-$z$ and $y$-$z$ planes (radian), 
	$\theta_{tx}^{\prime}$ and  $\theta_{ty}^{\prime}$ denote the UAV boresight angles (radian).

	Despite its simplicity, this model directly reflects UAV-induced effects such as jitter, wind deviations, and flight instability, which lead to misalignment and signal blockage. It is widely used in UAV-FSO research for its balance between interpretability and analytical tractability.

	\section{Adaptive Environment-Aware Transformer Autoencoder (AEAT-AE) Model Architecture}
	
	This section introduces the architecture of the proposed AEAT-AE model, including the latent representation design, the attention-based fusion of signal and environmental features, and the unified Transformer autoencoder for environment-aware encoding and reconstruction.
	
	\subsection{Latent Representation Design for Joint Signal and Environment Awareness}
	
	Before entering the Transformer layers, both the input signal and the environmental parameters are embedded into compatible latent spaces, enabling joint processing via attention mechanisms.

	\subsubsection{Tokenizing and Embedding Binary Signals for AEAT-AE Encoding}
	
	The original binary input $\mathbf{s} \in \{0,1\}^{N}$, where $N$ denotes the bitstream length, is first partitioned into a sequence of $T$ tokens, each containing $d_{\text{in}} = N/T$ bits,\footnote{For simplicity, $N/T$ is assumed to be an integer, which can be readily satisfied via appropriate framing or zero-padding in practical implementations.} 
	forming an initial token representation $\mathbf{X}_0 \in \mathbb{R}^{T \times d_{\text{in}}}$. Here, $T$ denotes the Transformer input sequence length.
	
	Each token is then linearly projected into the Transformer embedding space of dimension $d_k$ via
	\begin{equation}
		\mathbf{X} = \mathbf{X}_{0} W_{\text{sig}} + b_{\text{sig}}, 
		\quad W_{\text{sig}} \in \mathbb{R}^{d_{\text{in}} \times d_{k}}, \ 
		b_{\text{sig}} \in \mathbb{R}^{d_{k}},
	\end{equation}
	where $W_{\text{sig}}$ and $b_{\text{sig}}$ are the trainable projection matrix and bias vector, respectively. This tokenization and embedding define a deterministic mapping from $N$ input bits to a continuous representation of dimension $T \times d_k$, which is further reshaped into the channel input vector $\mathbf{x} \in \mathbb{R}^{K \times 1}$ with $K = T \cdot d_k$.

	\subsubsection{Processing and Embedding Environmental States for AEAT-AE Adaptation}
	
	The environment representation $\mathbf{E} = [Z, V_d, C_n^2, \sigma_s, \sigma_a]$ is designed to capture the key physical and operational factors affecting UAV–FSO signal propagation. Specifically, link distance $Z$ reflects path loss; visibility $V_d$ characterizes atmospheric attenuation; turbulence strength $C_n^2$ captures optical scintillation; and pointing error $\sigma_s$ along with UAV jitter $\sigma_a$ represent misalignment and platform instability. Together, these parameters provide a compact but informative summary of the environment to guide AEAT-AE adaptation.

	For practical implementation, the environmental parameters required by AEAT-AE can be estimated using commonly available UAV sensors. The link distance $Z$ can be obtained from onboard GPS/RTK or low-power LiDAR ranging. Visibility $V_d$ can be inferred from LiDAR return intensity, short-distance beacon probing, or onboard optical sensors, potentially combined with lightweight calibration or sensor fusion to improve accuracy. Turbulence strength $C_n^2$ can be approximated via scintillation indices, and pointing error $\sigma_s$ together with UAV jitter $\sigma_a$ can be derived from IMU statistics or optical centroid tracking. While sensor readings may contain noise or latency, lightweight smoothing or temporal aggregation can mitigate such effects. Since turbulence, visibility, and vibration parameters vary much slower than the FSO symbol rate, smoothed or slightly delayed measurements still provide meaningful context for AEAT-AE, enabling reliable DQN-based dynamic layer selection.
	
	Accordingly, $\mathbf{E}$ is designed to capture coarse yet informative environmental trends rather than precise, high-frequency reconstructions of atmospheric or mechanical dynamics. These trend-based features, derived from smoothed or aggregated measurements, still provide meaningful context for AEAT-AE adaptation. To align these heterogeneous features with the Transformer's input space, $\mathbf{E}$ is passed through the multi-layer perceptron (MLP):

	\begin{equation}
		\mathbf{E_{in}} = \sigma_E(\mathbf{E} W_{\text{env}} + b_{\text{env}}),
		\quad 
		W_{\text{env}} \in \mathbb{R}^{5 \times d_k},\;
		b_{\text{env}} \in \mathbb{R}^{d_k},
	\end{equation}
	
	\noindent where $\sigma_E(\cdot)$ is the nonlinear activation and $W_{\text{env}}, b_{\text{env}}$ are learnable parameters. The resulting vector $\mathbf{E_{in}} \in \mathbb{R}^{1 \times d_k}$ serves as the latent environmental token. 
	
	\subsection{Attention Mechanisms and Environment Fusion}
	
	Our proposed AEAT-AE employs Transformer's dual attention: self-attention for signal dependencies and cross-attention for environmental fusion, enabling dynamic feature adaptation to UAV-FSO channel variations.
	
	\begin{figure}[!t]
		\centering
		\includegraphics[width=2.7in]{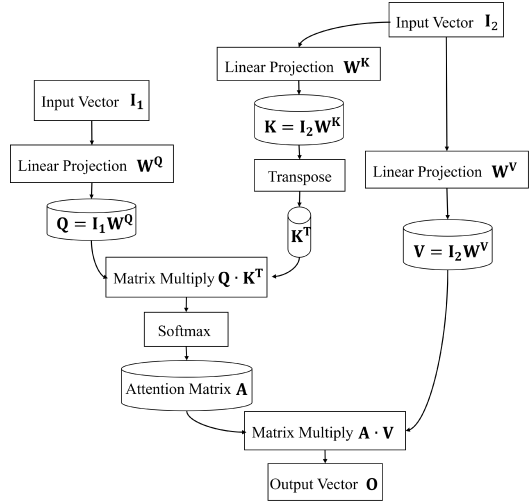}
		\caption{Illustration of the Attention Mechanism in Transformer Models.}
		\label{Attention}
	\end{figure}
	
	The resulting signal and environment embeddings and their interaction through attention are illustrated in Fig.~\ref{Attention}. The Transformer encoder consists of stacked layers with self-attention and cross-attention. The first layer takes input \(\mathbf{I}^{(1)} = \mathbf{X} \in \mathbb{R}^{T \times d_k}\). Subsequent layers use \(\mathbf{I}^{(l)}\), the output from the previous layer.
	
	\subsubsection{Signal Feature Extraction via Multi-Head Self-Attention}
	
	At the \(l\)-th layer, multi-head self-attention processes \(\mathbf{I}^{(l)} \in \mathbb{R}^{T \times d_k}\), capturing dependencies within the sequence. With \(N_h\) heads, the model attends to multiple representation subspaces simultaneously. For each head $i$, queries, keys, and values are computed as:
	
	\begin{equation}
		\begin{aligned}
			\mathbf{Q}_i^{(l)} &= \mathbf{I}^{(l)} \mathbf{W}_i^{Q,(l)}, \\
			\mathbf{K}_i^{(l)} &= \mathbf{I}^{(l)} \mathbf{W}_i^{K,(l)}, \\
			\mathbf{V}_i^{(l)} &= \mathbf{I}^{(l)} \mathbf{W}_i^{V,(l)}, 
		\end{aligned}
	\end{equation}
	
	\noindent where \(\mathbf{W}_i^{Q,(l)}, \mathbf{W}_i^{K,(l)} \in \mathbb{R}^{d_k \times d_k}\) and \(\mathbf{W}_i^{V,(l)} \in \mathbb{R}^{d_k \times d_v}\) are learnable parameters. $d_v=d_k/N_h$ denotes the feature dimension of each head’s value projection. The attention weights are computed using scaled dot-product attention:
	
	\begin{equation}
		\mathbf{A}_i^{(l)} = \text{Softmax}\left(\frac{\mathbf{Q}_i^{(l)} (\mathbf{K}_i^{(l)})^\top}{\sqrt{d_k}}\right),
		\label{Att}
	\end{equation}

	\noindent where the Softmax is applied row-wise to normalize attention scores for each query token. The scaling factor $\sqrt{d_k}$ controls the magnitude of the dot-product in the numerator,   preventing overly large values from pushing the Softmax into saturation and thereby ensuring numerically stable gradients during training.
	
	Then, it is followed by the head-wise output:
	
	\begin{equation}
		\mathbf{O}_i^{(l)} = \mathbf{A}_i^{(l)} \mathbf{V}_i^{(l)}.
		\label{Ott}
	\end{equation}
	
	The outputs from all heads are concatenated and linearly transformed to produce the self-attention output:
	
	\begin{equation}
		\mathbf{O}_\text{{self}}^{(l)} = \text{Concat}(\mathbf{O}_1^{(l)}, \ldots, \mathbf{O}_{N_h}^{(l)}) \mathbf{W}^{O,(l)},
	\end{equation}
	
	\noindent where \(\mathbf{W}^{O,(l)} \in \mathbb{R}^{N_h d_v \times d_k}\) is the output projection matrix.

	\subsubsection{Environmental Feature Fusion via Cross-Attention} 
	To incorporate environmental information, each layer includes a cross-attention module. The self-attention output \(\mathbf{O}_{\text{self}}^{(l)}\) is first combined with the layer input \(\mathbf{I}^{(l)}\) via a residual connection, followed by layer normalization to produce \(\mathbf{N}_{\text{self}}^{(l)}\), which serves as the query in the cross-attention. 
	
	The environmental embedding $\mathbf{E_{in}} \in \mathbb{R}^{1 \times d_k}$ is treated as a single global token. In cross-attention, each signal token in $\mathbf{N}_{\text{self}}^{(l)}$ serves as a query to attend to this single environmental token. This design avoids any temporal tiling or duplication, maintaining minimal redundancy and consistent environment context for all signal tokens.

	The attention mechanism in this cross-attention module is similar to that in self-attention, where the query \(\mathbf{Q}_i^{c,(l)}\), key \(\mathbf{K}_i^{c,(l)}\), and value \(\mathbf{V}_i^{c,(l)}\) are computed as:
	\begin{equation}
		\begin{aligned}
			\mathbf{Q}_i^{c,(l)} &= \mathbf{N}_{\text{self}}^{(l)} \mathbf{W}_i^{Q_c,(l)}, \\
			\mathbf{K}_i^{c,(l)} &= \mathbf{E_{in}} \mathbf{W}_i^{K_c,(l)}, \\
			\mathbf{V}_i^{c,(l)} &= \mathbf{E_{in}} \mathbf{W}_i^{V_c,(l)}.
		\end{aligned}
	\end{equation}
	Here, $\mathbf{W}_i^{Q_c,(l)}, \mathbf{W}_i^{K_c,(l)} \in \mathbb{R}^{d_k \times d_k}$ and $\mathbf{W}_i^{V_c,(l)} \in \mathbb{R}^{d_k \times d_v}$ are learnable projection matrices ensuring that the dimensions of queries ($T\times d_k$), keys ($1\times d_k$), and values ($1\times d_v$) are consistent for the subsequent scaled dot-product attention.
	
	The attention weights are computed through scaled dot-product attention similar to the operations in (\ref{Att}) and (\ref{Ott}), with the output of $\mathbf{O}_{i}^{c,(l)}$. The final fused output \(\mathbf{O}_\text{{cross}}^{(l)}\) is obtained by concatenating the outputs of all attention heads and applying a linear transformation:
	\begin{equation}
		\mathbf{O}_\text{{cross}}^{(l)} = \text{Concat}(\mathbf{O}_1^{c,(l)}, \ldots, \mathbf{O}_{N_h}^{c,(l)}) \mathbf{W}^{O_c,(l)}.
	\end{equation}
	
	This fused output \(\mathbf{O}_\text{{cross}}^{(l)}\) incorporates the environmental information to adjust the self-attention output, allowing the model to jointly attend to both the signal and environmental contexts.

	\subsection{Transformer-Based Autoencoder in AEAT-AE}
	
	\begin{figure*}[!t] 
		\centering 
		\includegraphics[width=6in]{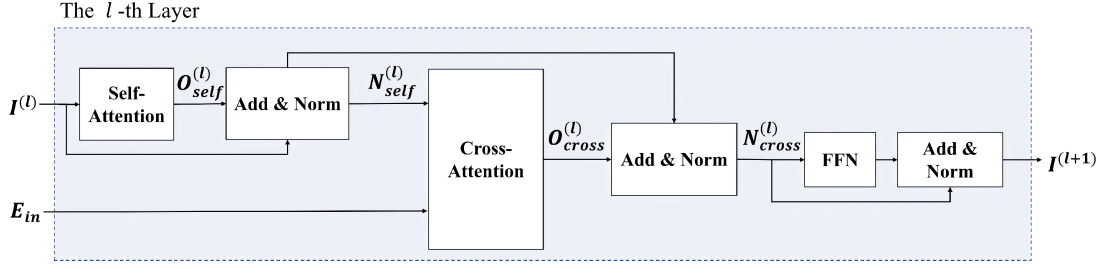} 
		\caption{Unified Transformer Layer.}
		\label{Layer} 
	\end{figure*}
	
	\subsubsection{Unified Transformer Layer Design}
	
	The input embedding and attention mechanism form the foundational components of our proposed AEAT-AE, which employs a Transformer-AE as its backbone architecture. A key feature of this design is the shared Transformer layer structure between encoder and decoder (Fig.~\ref{Layer}), where each layer integrates the following modules:
	
	\begin{itemize}
		\item {Self-Attention Module:} Captures intra-modal dependencies within the signal embedding.
		\item {Cross-Attention Module:} Integrates environmental context into the signal representation.
		\item {Feedforward Network (FFN) and Layer Normalization:} Enhances feature transformation and stabilizes training.
	\end{itemize}
	
	\paragraph{Self-Attention Module}
	
	The self-attention mechanism refines the signal representation by capturing intra-modal relationships. The output of multi-head self-attention is:
	\begin{equation}
		\mathbf{O}_\text{self}^{(l)} = \text{SelfAttention}(\mathbf{I}^{(l)}),
	\end{equation}
	\noindent where $\mathbf{O}_\text{self}^{(l)} \in \mathbb{R}^{T \times d_k}$ is the self-attended signal feature.
	
	\paragraph{Cross-Attention Module}
	
	{Unlike standard Transformer encoders that rely solely on self-attention, AEAT-AE applies cross-attention at every encoder and decoder layer to incorporate environmental parameters. Since environmental states evolve much more slowly than per-symbol features, reusing the same environment token at each layer preserves these priors throughout the network, preventing them from being diluted by residual connections or deep nonlinear transformations.}
	
	To fuse external environmental information, we apply cross-attention between the self-attended output $\mathbf{O}_\text{self}^{(l)}$ and the embedded environmental sequence $\mathbf{E}_{\text{in}}$:
	\begin{equation}
		\mathbf{O}_\text{cross}^{(l)} = \text{CrossAttention}(\mathbf{O}_\text{self}^{(l)}, \mathbf{E}_{\text{in}}),
	\end{equation}
	\noindent where $\mathbf{O}_\text{cross}^{(l)} \in \mathbb{R}^{T \times d_k}$ produces environment-conditioned features for downstream processing.
	
	{
		By applying cross-attention at every layer, the encoder continuously aligns the evolving signal features with the environmental embedding.  
		This layer-wise conditioning mitigates the risk of early-stage environmental cues being diluted by residual connections or deep nonlinear transformations,  
		thereby ensuring that channel-related information remains influential throughout the entire encoding process.}

	\paragraph{Feed-Forward Network and Layer Normalization}
	
	The output of each attention module undergoes a series of standard Transformer operations: residual connection, layer normalization, and position-wise FFN.
	
	First, the self-attention output $\mathbf{O}_{\text{self}}^{(l)}$ is combined with the layer input $\mathbf{I}^{(l)}$ through residual connection and layer normalization:
	\begin{equation}
		\mathbf{N}_{\text{self}}^{(l)} = \text{LayerNorm}\left(\mathbf{I}^{(l)} + \mathbf{O}_{\text{self}}^{(l)}\right).
	\end{equation}
	
	Subsequently, the cross-attention output $\mathbf{O}_{\text{cross}}^{(l)}$ is integrated using the same residual architecture:
	\begin{equation}
		\mathbf{N}_{\text{cross}}^{(l)} = \text{LayerNorm}\left(\mathbf{N}_{\text{self}}^{(l)} + \mathbf{O}_{\text{cross}}^{(l)}\right).
	\end{equation}
	
	The normalized output is then processed by a position-wise feed-forward network:
	\begin{equation}
		\mathbf{F}^{(l)} = \text{FFN}\left(\mathbf{N}_{\text{cross}}^{(l)}\right),
	\end{equation}
	\noindent where the $\text{FFN}(\cdot)$ comprises two linear transformations with ReLU activation.
	
	The final output of the $l$-th layer is obtained through the last residual connection and normalization, and then serves as the input to the subsequent Transformer layer ($l+1$):
	
	\begin{equation}
		\mathbf{I}^{(l+1)} = \text{LayerNorm}\left(\mathbf{N}_{\text{cross}}^{(l)} + \mathbf{F}^{(l)}\right).
	\end{equation}
	
	{
		Finally, by deploying this unified framework symmetrically in both encoder and decoder, AEAT-AE ensures that both latent compression and reconstruction remain jointly aware of channel conditions. This symmetric design prevents information mismatch between the two stages and improves robustness under highly dynamic UAV-FSO environments.}

	\subsubsection{AEAT-AE Encoder: Environment-Aware Representation Learning}
	
	The goal of the Transformer encoder is to produce a latent representation that captures both the signal structure and the impact of environmental parameters. For each Transformer layer $l \in \{1, \dots, L\}$, the signal representation is updated as:
	\begin{equation}
		\mathbf{I}^{(l)} = \mathrm{TransformerLayer}(\mathbf{I}^{(l-1)}, \mathbf{E_{in}}), \quad \mathbf{I}^{(0)} = \mathbf{X}.
	\end{equation}
	\noindent where $\mathbf{I}^{(l)} \in \mathbb{R}^{T \times d_k}$ contains encoded features at layer $l$. The final output $\mathbf{I}^{(L)}$ integrates signal dependencies conditioned on $\mathbf{E}_{\text{in}}$. {Before transmission over the UAV-FSO channel, $\mathbf{I}^{(L)}$ is further processed via a linear projection and physical-layer constraints to generate the transmit vector $\mathbf{x} \in \mathbb{R}^{K \times 1}$ compatible with IM/DD optical modulation.}

	\subsubsection{AEAT-AE Decoder: Signal Recovery under Environmental Conditions}
	
	The decoder aims to reconstruct the original signal using the received distorted signal $\mathbf{y} \in \mathbb{R}^{K \times 1}$ and the same environmental parameters $\mathbf{E_{in}}$. 
	
	To align with the Transformer's expected input, $\mathbf{y}$ is projected into the embedding space using the same projection strategy as the encoder:
	\begin{equation}
		\mathbf{R}^{(0)} = \mathrm{Proj}(\mathbf{y}) \in \mathbb{R}^{T \times d_k}.
	\end{equation}
	
	Each decoder layer then updates the representation as:
	\begin{equation}
		\mathbf{R}^{(l)} = \mathrm{TransformerLayer}(\mathbf{R}^{(l-1)}, \mathbf{E_{in}}), \quad l = 1, \dots, L.
	\end{equation}
	
	\noindent Here, $\mathbf{R}^{(l)} \in \mathbb{R}^{T \times d_k}$ denotes the output of layer $l$. After $L$ decoding layers, the final output is projected back to the signal domain:
	\begin{equation}
		\hat{\mathbf{X}} = \mathrm{Linear}(\mathbf{R}^{(L)}),
	\end{equation}
	\noindent where the linear layer restores the original signal dimensionality.
	
	{While the full-depth Transformer-AE provides strong reconstruction capability, its fixed computational footprint prevents efficient adaptation to the highly dynamic UAV–FSO environment. As channel conditions may vary dramatically across altitude, turbulence, and pointing stability, a mechanism that allows the autoencoder to adjust its effective depth in real time becomes essential. This motivates the adaptive layer control introduced in the next section.}

	\section{{Adaptive Layer Control and System Workflow}}

	{To mitigate the computational cost of deep Transformer stacks while adapting to dynamic UAV-FSO channels, the AEAT-AE incorporates a lightweight controller that selectively activates Transformer layers based on the current channel conditions. Simple rule-based strategies are inadequate due to the coupling of multiple environmental factors, and tabular Q-learning is impractical for continuous channel states. Instead, a DQN-based controller directly learns adaptive layer policies from interactions with the environment, enabling efficient, real-time selection of relevant layers without manual thresholds or state discretization.}

	\subsection{Adaptive Layer Selection via DQN}
	
	The DQN agent determines which Transformer layers should be activated based on the current channel condition. The following components describe the RL formulation tailored for AEAT-AE.
	
	{State (\(S\))}: The state \(S\) encodes the current environment as a continuous vector \(\mathbf{E} = [Z, V_d, C_n^2, \sigma_s, \sigma_a]\), capturing diverse atmospheric and alignment factors. Prior to input into the DQN, this vector is normalized to ensure numerical stability and balanced feature scaling. Avoiding discretization preserves fine-grained variations, allowing the DQN to learn smooth, adaptive layer selection policies under dynamic channel conditions.

	{Action space (\(\mathcal{A}\)): The action space (\(\mathcal{A}\)) consists of all possible activation patterns of the $L$ Transformer layers. Each action is encoded as an integer \(a \in \{1, \dots, 2^L - 1\}\), whose binary form specifies the active layers. Based on preliminary experiments and practical considerations, \(L\) is typically kept below 4, keeping the action space manageable. The same $L$-layer Transformer template is used for both encoder and decoder, and inactive layers (bit 0) are replaced with an identity mapping plus LayerNorm to preserve feature scale and avoid distribution drift. This design enables efficient, stable, and adaptive layer selection in real time.}

	Given an action $a$, the number of activated layers $L_a$ can be presented as:
	
	\begin{equation}
		L_a = \mathrm{HammingWeight}(a),
	\end{equation}
	\noindent which counts the number of bits equal to one in the binary expansion of $a$. This provides a precise, architecture-independent measure of computational cost for use in the reward function. To make this quantity comparable with the normalized bit error rate in the reward function, we define the normalized activated layer ratio as:
	\begin{equation}
		\tilde{L}_a = \frac{L_a}{L} \in [0,1].
	\end{equation}
	
	{Reward (\(r\))}: The immediate reward balances communication quality and computational cost, defined as:
	
	\begin{equation}
		r = -\left(\lambda_{r_1} \cdot \tilde{\mathrm{BER}} + \lambda_{r_2} \tilde{L}_a \right),
		\label{reward}
	\end{equation}
	\noindent where \(\lambda_{r_1}, \lambda_{r_2}\) regulate the trade-off between BER and the number of activated layers, \(\tilde{\mathrm{BER}} = \min(-\log_{10}(\mathrm{BER}), 10)/10 \in [0,1]\) is the normalized BER. This ensures that both terms contribute comparably to the reward and avoids scale mismatch.
	
	{The weights were set to a balanced ratio \(\lambda_{r_1} : \lambda_{r_2} = 1 : 1\), providing stable training and consistent policy behavior. These coefficients can be adjusted in future studies to trade off reliability, complexity, and latency.}

	{Q-value Approximation and Learning}: The agent employs a neural network parameterized by weights \(\theta\) to approximate the action-value function \(Q(S,a;\theta)\). The network is trained by minimizing the temporal difference loss
	
	\begin{equation}
		\mathcal{L}(\theta) = \mathbb{E}_{(S,a,r,S')} \left[ \left( r + \gamma \max_{a'} Q(S', a'; \theta^-) - Q(S,a; \theta) \right)^2 \right],
		\label{loss}
	\end{equation}
	
	\noindent where the tuple \((S,a,r,S')\) consists of the current state \(S\), action \(a\), received reward \(r\), and the subsequent state \(S'\). The variable \(a'\) denotes possible next actions in the next state \(S'\). Here, \(\theta^-\) refers to the parameters of a periodically updated target network that stabilizes training, and \(\gamma\) is the discount factor reflecting the importance of future rewards.
	
	{Experience Replay}: Transitions \((S,a,r,S')\) are stored in a prioritized replay buffer \(\mathcal{D}\). Sampling based on temporal difference errors focuses training on more informative experiences, enhancing sample efficiency and convergence speed.
	
	{Policy Update and Inference}: The DQN agent updates \(\theta\) iteratively via gradient descent on \(\mathcal{L}(\theta)\). During inference, given the current state \(S\), the optimal action \(a^*\) is selected as
	
	\begin{equation}
		a^* = \arg\max_{a} Q(S,a;\theta).
	\end{equation}
	
	By adaptively selecting the most informative Transformer layers based on environmental conditions, the system reduces computation and overfitting while ensuring robust reconstruction. The DQN's function approximation efficiently handles high-dimensional states, enabling real-time, data-driven optimization beyond heuristic or fixed rules.

	\subsection{Training and Deployment Process of the DQN Controller} 
	
	With the state, action, and reward definitions established above, the DQN controller can now be trained to learn optimal layer-selection policies. This training enables the AEAT-AE system to adapt its Transformer depth dynamically under varying UAV-FSO channel conditions. Training and deployment proceed in two stages.

	At each training step, given the current state \(S\), the controller selects an action \(a\) using an \(\varepsilon\)-greedy policy, where \(\varepsilon\) controls the exploration probability:
	
	\begin{equation}
		a =
		\begin{cases}
			\text{random action}, & \text{with probability } \varepsilon, \\
			\arg\max_{a'} Q(S, a'; \theta), & \text{with probability } 1 - \varepsilon.
		\end{cases}
	\end{equation}
	
	After executing the selected Transformer layers, the system evaluates the resulting performance, receives a reward \(r\), observes the new state \(S'\), and stores the transition \((S, a, r, S')\) in a (prioritized) replay buffer for sample-efficient training. The network parameters \(\theta\) are updated by minimizing the temporal difference loss as defined in (\ref{loss}), with a periodically updated target network \(\theta^-\) and discount factor \(\gamma\). These parameter updates are performed during the training phase, while the learned policy is fixed during deployment.
	
	During deployment, when a reconfiguration decision is triggered, the system selects the optimal action \(a^* = \arg\max_a Q(S, a; \theta)\) for a given state \(S\) without exploration, enabling real-time, data-driven layer selection.
	
	Although the DQN controller is lightweight and capable of real-time layer selection, frequent reconfigurations can degrade system stability. To mitigate this, a change-detection mechanism is incorporated, which updates the Transformer configuration only when the system state exhibits significant variation. Specifically, let \(S_t\) and \(S_{t-1}\) denote the normalized state vectors at two consecutive decision instants. The magnitude of state variation is then quantified as:
	
	\begin{equation}
		\Delta S_t = \left\| W_s \cdot (S_t - S_{t-1}) \right\|_2,
	\end{equation}
	\noindent where \(W_s\) is a diagonal weighting matrix used to account for the relative scales and sensitivities of different state components.
	
	If \(\Delta S_t\) exceeds a predefined threshold \(\delta\), a new DQN action is triggered to update the Transformer layer configuration; otherwise, the current configuration is retained. This mechanism leverages the DQN's generalization to ignore minor state fluctuations while responding to significant environmental changes, effectively balancing adaptability and computational efficiency. The threshold \(\delta\) is treated as a system-level design parameter that can be selected offline to trade off reconfiguration frequency and computational overhead according to task requirements and available computing resources, while remaining fixed during deployment.

	\subsection{Overall AEAT-AE Structure}
	
	To implement the adaptive layer selection described above, the AEAT-AE framework integrates the Transformer-AE and the DQN-based controller toward a unified system-level objective of achieving high reconstruction fidelity with efficient layer utilization. Algorithm~\ref{alg:training_deployment_dqn} outlines the overall procedure, including offline training of the Transformer-AE, subsequent DQN policy learning with fixed AE parameters, and online deployment. This sequential training strategy ensures stable reconstruction behavior, avoids reward drift during policy learning, and facilitates reproducible and reliable system optimization.
	
	\begin{algorithm}[ht] 
		\caption{Training and Deployment of AEAT-AE with DQN-Based Adaptive Layer Selection}
		\label{alg:training_deployment_dqn}
		\small
		\begin{algorithmic}[1]
			\Require Input binary signal $\mathbf{s}$; environment parameters $\mathbf{E} = [Z, V_d, C_n^2, \sigma_s, \sigma_a]$; channel coefficient $h$ (scalar, only for simulation); AWGN noise $\mathbf{w}$
			\Ensure Reconstructed signal $\hat{\mathbf{s}}$
			
			\Statex \textbf{Step 1: Full-depth Transformer-AE pre-training (all layers active)}
			\State Define subroutine: $\hat{\mathbf{s}} = \text{AEAT\_AE}(\mathbf{s}, \mathbf{E}, h, a)$
			\State Embed input: $\mathbf{X} = \text{Embed}(\mathbf{s})$
			\State Encode: $\mathbf{x} = \text{Encoder}(\mathbf{X}; \theta_{\text{enc}}, a)$
			\State Channel transmission: $\mathbf{y} = h \cdot \mathbf{x} + \mathbf{w}$
			\State Project received signal: $\mathbf{R}^{(0)} = \mathrm{Proj}(\mathbf{y}) \in \mathbb{R}^{T \times d_k}$
			\State Decode: $\hat{\mathbf{s}} = \text{Decoder}(\mathbf{R}^{(0)}; \theta_{\text{dec}}, a)$
			\State Compute BCE loss: 
			\[
			\mathcal{L}_{\mathrm{AE}} = -\sum_{n=1}^{N} \left( s_n \log(\hat{s}_n) + (1 - s_n) \log(1 - \hat{s}_n) \right)
			\]
			\State Update AE parameters: 
			\[
			\theta_{\mathrm{AE}} \leftarrow \arg\min \lambda_1 \mathcal{L}_{\mathrm{AE}} + \lambda_4 \|\theta_{\mathrm{AE}}\|^2
			\]
			
			\Statex \textbf{Step 2: DQN-based adaptive layer selection (policy learning)}
			\State Freeze $\theta_{\mathrm{AE}}$; initialize DQN $Q(S_t,a_t;\theta)$ and target network $Q(S_t,a_t;\theta^-)$
			\State Initialize replay buffer $\mathcal{D}$
			\For{each training episode}
			\State Sample environment parameters $\mathbf{E}$ and normalize as RL state $S_t$
			\State Select action $a_t$ (subset of layers) using $\varepsilon$-greedy policy over $Q(S_t, a; \theta)$
			\State Reconstruct signal: $\hat{\mathbf{s}} = \text{AEAT\_AE}(\mathbf{s}, \mathbf{E}, h, a_t)$
			
			\State Compute BER: $\mathrm{BER} = \frac{1}{N}\sum_{n=1}^{N} \mathbb{1}(\hat{s}_n \ne s_n)$
			
			\State Count activated layers: $L_{a_t} = \mathrm{HammingWeight}(a_t)$
			
			\State Compute reward: 
			\[
			r_t = -(\lambda_{r_1} \cdot \tilde{\mathrm{BER}} + \lambda_{r_2} \cdot \tilde{L}_{a_t})
			\]
			
			\State Observe next state $S_{t+1}$
			\State Store transition $(S_t, a_t, r_t, S_{t+1})$ in buffer $\mathcal{D}$
			
			\State Sample mini-batch $(S_i, a_i, r_i, S_i')$ from $\mathcal{D}$
			\State Compute TD target: 
			\[
			y_i = r_i + \gamma \max_{a'} Q(S_i', a'; \theta^-)
			\]
			\State Update DQN by minimizing TD loss:
			\[
			\min_{\theta}\ \mathcal{L}_{\mathrm{DQN}} = \frac{1}{B} \sum_i \left( y_i - Q(S_i, a_i; \theta) \right)^2
			\]
			\State Periodically update target network: $\theta^- \leftarrow \theta$
			\EndFor
			
			\Statex \textbf{Step 3: AEAT-AE deployment (real-world, channel unknown)}
			\State Observe current environment parameters $\mathbf{E}$ and normalize as RL state $S_t$
			\State Select optimal action: $a^* = \arg\max_a Q(S_t, a; \theta)$
			\State Reconstruct signal: $\hat{\mathbf{s}} = \text{AEAT\_AE}(\mathbf{s}, \mathbf{E}, a^*)$
			\State If significant change in $\mathbf{E}$ is detected, remap to $S_{t+1}$ and reselect action $a^*$
		\end{algorithmic}
	\end{algorithm}

	\subsubsection{Training Process}
	
	\begin{figure*}[!t]
		\centering
		\includegraphics[width=7in]{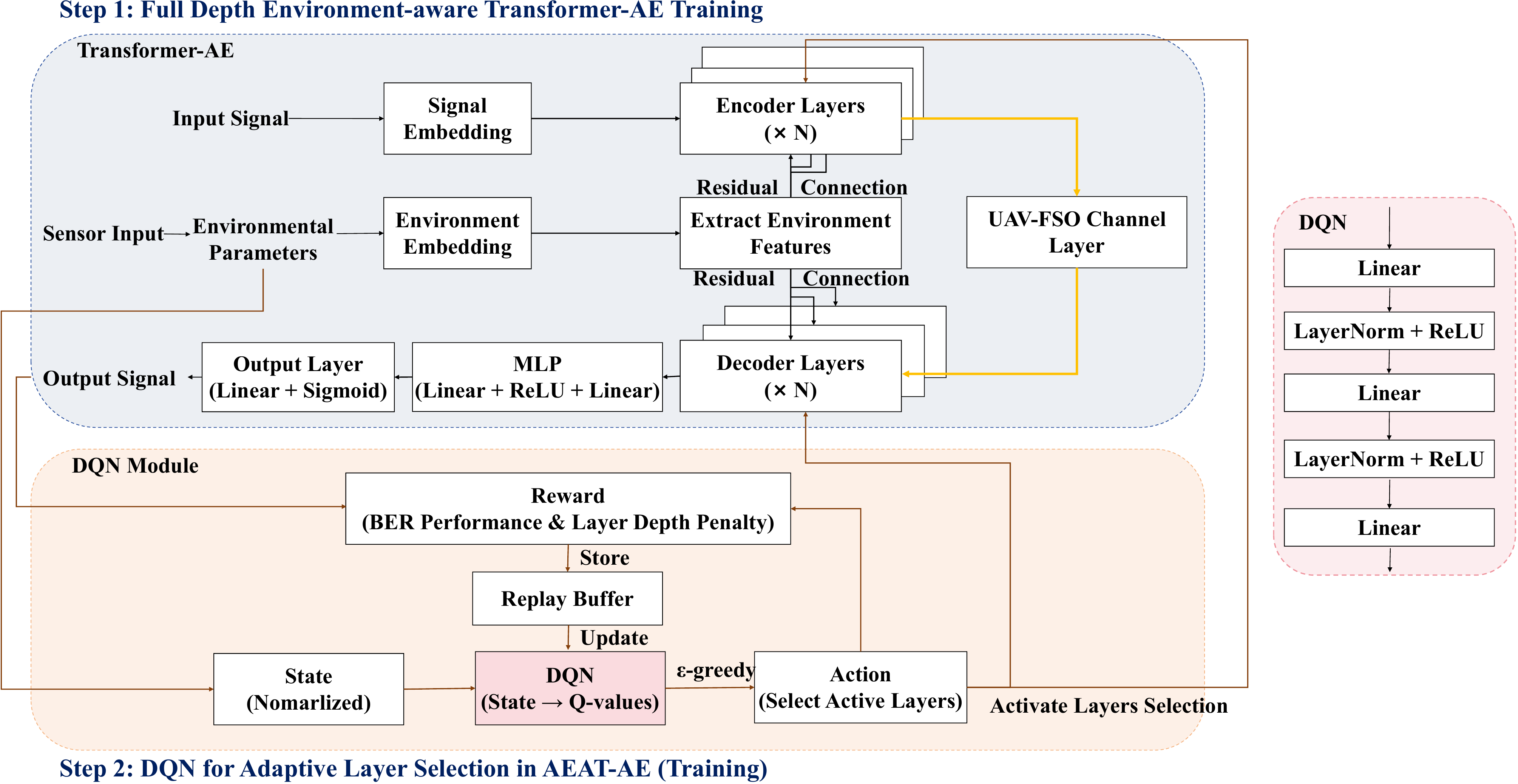}
		\caption{Training process of AEAT-AE.}
		\label{Training}
	\end{figure*}
	
	As illustrated in Fig.~\ref{Training} and Algorithm~\ref{alg:training_deployment_dqn}, the AEAT-AE framework is trained in two sequential stages, corresponding to Transformer-AE pre-training and DQN-based adaptive policy learning.
	
	In the first stage, the Transformer-AE is trained with all $L$ layers activated to establish a robust baseline for signal reconstruction under diverse channel conditions. As discussed in Sec.~II, the autoencoder is optimized using the BCE loss:
	
	\begin{equation}
		\mathcal{L}_{\mathrm{AE}}(\mathbf{s}, \hat{\mathbf{s}}) = -\sum_{n=1}^{N} \left( s_n \log(\hat{s}_n) + (1 - s_n) \log(1 - \hat{s}_n) \right),
	\end{equation}
	\noindent where BCE is adopted because it is fully differentiable and enables stable end-to-end gradient-based training of the Transformer-AE. The optimization objective in this stage is given by:
	
	\begin{equation}
		\min_{\theta_{\mathrm{AE}}} \ \lambda_1 \, \mathcal{L}_{\mathrm{AE}} + \lambda_4 \|\theta_{\mathrm{AE}}\|^2,
	\end{equation}
	where $\theta_{\mathrm{AE}} = \theta_{\text{enc}} \cup \theta_{\text{dec}}$ denotes the encoder and decoder parameters, and $\lambda_1$ and $\lambda_4$ are balancing coefficients.
	
	In the second stage, the Transformer-AE parameters are fixed, and a DQN agent is trained to learn an adaptive Transformer layer selection policy. Following the DQN formulation in Sec.IV-A, the reward defined in (\ref{reward}) is directly used for policy learning during the second stage, without backpropagation through the Transformer-AE. 
	
	A key design choice in this stage is to use the BER directly as the reward. Unlike AE pre-training, DQN does not require gradient backpropagation through the reward, and using BER aligns the reward with the actual system-level performance while providing task-relevant feedback for layer-selection decisions, since the effect of activating or skipping Transformer layers is ultimately reflected in discrete decoding errors rather than continuous reconstruction deviations.
	
	Then the DQN is trained by minimizing the temporal-difference loss:
	
	\begin{equation}
		\mathcal{L}_{\mathrm{DQN}}(\theta) = \frac{1}{B} \sum_i \left( y_i - Q(S_i, a_i; \theta) \right)^2,
	\end{equation}
	with the TD target
	\begin{equation}
		y_t = r_t + \gamma \max_{a'} Q(S_{t+1}, a'; \theta^-),
	\end{equation}
	where $\theta^-$ denotes the parameters of the target network, $\gamma$ is the discount factor, and $S_t$ represents the continuous state constructed from the normalized environmental vector.
	
	For completeness, we present a conceptual system-level objective that unifies the goals of the two training stages. This formulation serves as a theoretical description rather than an instruction for joint parameter updates:
	\begin{align}
		\min_{\theta_{\mathrm{AE}}, \theta} \ 
		\mathbb{E}_{\mathbf{E}, a_t, \mathbf{s}} \big[ & 
		\lambda_1 \, \mathcal{L}_{\mathrm{AE}}(\mathbf{s}, \hat{\mathbf{s}})
		+ \lambda_2 \, (Q(S_t, a_t; \theta) - y_t)^2 \nonumber \\
		& + \lambda_3 \|\theta\|^2 + \lambda_4 \|\theta_{\mathrm{AE}}\|^2
		\big],
	\end{align}
	subject to
	\begin{equation}
		\left\{
		\begin{aligned}
			& L_{\min} \leq L_{a_t} \leq L, \\
			& a_t \in \mathcal{A}, \\
			& a_t = \arg\max_{a} Q(S_t, a; \theta),
		\end{aligned}
		\right.
	\end{equation}
	which enforces valid action selection and a minimum model capacity during policy learning. Here, $\lambda_1$ and $\lambda_2$ weight the reconstruction loss of the Transformer-AE and the temporal-difference loss of the DQN, respectively, while $\lambda_3$ and $\lambda_4$ control $\ell_2$ regularization on the DQN and AE parameters. We emphasize that these coefficients are not used for simultaneous optimization; instead, they conceptually reflect the relative contributions of the two training stages, which are carried out sequentially in practice as described above.
	
	Through iterative interaction with diverse simulated channel environments, the DQN progressively learns to activate only the necessary Transformer layers, achieving an effective accuracy–complexity trade-off under varying channel conditions.

	\subsubsection{Deployment Process}
	
	During deployment, the trained AEAT-AE system operates with fixed parameters. It dynamically selects Transformer layers based on real-time environmental states using the learned DQN policy.
	
	The deployment workflow, as detailed in Algorithm~\ref{alg:training_deployment_dqn}, is as follows: Observe current environmental parameters $\mathbf{E} = [Z, V_d, C_n^2, \sigma_s, \sigma_a]$ and normalize them to form the state $S_t$. The trained Q-network selects the optimal action:
	\begin{equation}
		a^* = \arg\max_a Q(S_t, a; \theta),
	\end{equation}
	which specifies the subset of Transformer layers to activate. The input signal is processed through the dynamically configured AEAT-AE, with inactive layers bypassed using identity mappings.
	
	To prevent excessive reconfigurations, policy recomputation occurs only when the L2-norm of the state change exceeds a threshold \(\delta\), filtering high-frequency fluctuations while responding to meaningful environmental variations. Computational complexity scales with $O(L_{\text{active}} \cdot (T^2 d_k + T d_k^2))$, where $L_{\text{active}}$ is the number of active layers, $T$ the Transformer input length, and $d_k$ the embedding dimension. DQN evaluation adds negligible overhead ($O(1)$), and infrequent policy updates ensure stable reconstruction with substantial computational savings.
	

	\section{Experiments Results}
	
	\subsection{Experiment Setup}
	
	During training, the UAV–FSO channel is simulated in MATLAB to generate the corresponding fading coefficients, which are then embedded into the channel layer of the proposed model. The main environmental and channel parameters used for this simulation are listed in Table~\ref{channel settings}. To improve generalization, each environmental vector is defined as $\mathbf{E} = [Z, V_d, C_n^2, \sigma_s, \sigma_a]$, and all elements are randomly sampled within their respective ranges. A similar training strategy has been adopted in many deep-learning-based communication studies, where a fixed mid-range SNR is used to stabilize optimization and prevent the loss surface from changing with noise power \cite{Related1, bourtsoulatze2019deep}. Training across multiple SNRs introduces a continuously shifting input distribution, making convergence more difficult. Using a representative SNR of 10 dB therefore provides a stable training condition while still allowing the model to generalize to unseen SNR levels, as confirmed in the evaluation results.

	For evaluation, the trained model is tested under a range of SNR values to assess its robustness to channel degradation. The test set is generated using the same parameter ranges as in training but without any overlap to prevent data leakage. BER is obtained through large-scale Monte Carlo simulations.
	
	The inference time of the trained model was measured on an Intel Xeon W5-3435X CPU and an NVIDIA RTX 4090 GPU. The reported times reflect end-to-end forward passes for the test samples, providing a reference for the computational efficiency of the proposed system. Table~\ref{network_settings} summarizes the key hyperparameters used in the AEAT-AE framework, covering both the Transformer-AE and the DQN-based adaptive layer selection.
	
	\begin{table}[t]
		\centering
		\caption{Channel Simulation Variables Settings}
		\begin{tabular}{|>{\centering\arraybackslash}m{1.0cm}|
				>{\centering\arraybackslash}m{2.6cm}|
				>{\centering\arraybackslash}m{1.1cm}|
				>{\centering\arraybackslash}m{2.2cm}|}
			\hline
			\textbf{Variable} & \textbf{Value} & \textbf{Variable} & \textbf{Value} \\ \hline
			$\sigma_s$ & $0.01 \le \sigma_s \le 0.05$ m 
			& $\rho$ & 0.596 \\ \hline
			$\eta_e$ & 0.5 
			& $\phi_A - \phi_B$ & $\pi/2$ \\ \hline
			$\sigma_a$ & $2 \le \sigma_a \le 5$ mrad 
			& $\Omega$ & 1.3265 \\ \hline
			$\sigma_w^2$ & $10^{-10}$ 
			& $b_0$ & 0.1079 \\ \hline
			$\lambda$ & 1550 nm 
			& $\alpha$ & 8.2 \\ \hline
			$\beta$ & 4 
			& $r_a$ & 0.1 m \\ \hline
			$w_z$ & 1 m
			& $\theta_{\text{FoV}}$ & 20 mrad \\ \hline
			$V_d$ & $2 \le V_d \le 13$ km 
			& $Z$ & $1 \le Z \le 5$ km \\ \hline
			\multicolumn{4}{|c|}{$C_n^2 = \{5{\times}10^{-14}, 1.7{\times}10^{-14}, 5{\times}10^{-15}, 4{\times}10^{-15}\}$} \\ \hline
		\end{tabular}
		\label{channel settings}
	\end{table}

	\begin{table}[h]
		\centering
		\caption{Parameter Settings for AEAT-AE}
		\begin{tabularx}{\linewidth}{|>{\raggedright\arraybackslash}m{6cm}|>{\centering\arraybackslash}X|}
			\hline
			\multicolumn{2}{|c|}{\textbf{Transformer-AE Settings}} \\
			\hline
			Sequence length $T$ & 16 \\
			\hline
			Embedding dimension & 32 \\ 
			\hline
			Number of attention heads $N_h$ & 4 \\
			\hline
			Number of Transformer layers $L$ & 4 \\
			\hline
			Initial active layers & [0, 1, 2, 3] \\
			\hline
			Environmental parameter dimension & 5 \\
			\hline
			Batch size & 64 \\
			\hline
			Transformer-AE training epochs & 50 \\
			\hline
			Learning rate & 0.001 \\
			\hline
			Fixed training SNR & 10 dB \\
			\hline
			\multicolumn{2}{|c|}{\textbf{DQN Settings}} \\
			\hline
			State dimension & 5 \\
			\hline
			Learning rate & $3 \times 10^{-4}$ \\
			\hline
			Discount factor $\gamma$ & 0.99 \\
			\hline
			Soft target update rate $\tau$ & $1 \times 10^{-3}$ \\
			\hline
			Initial $\epsilon$ & 1.0 \\
			\hline
			Minimum $\epsilon$ & 0.01 \\
			\hline
			$\epsilon$ decay rate & 0.995 \\
			\hline
			Replay memory size & 100000 \\
			\hline
			Batch size & 128 \\
			\hline
			Prioritized replay alpha & 0.6 \\
			\hline
			Prioritized replay beta & 0.4 \\
			\hline
			Optimizer & AdamW \\
			\hline
			Learning rate scheduler & CosineAnnealingLR \\
			\hline
		\end{tabularx}
		\label{network_settings}
	\end{table}

	\subsection{Performance Results}
	
	To analyze the contribution of each component in our framework, we construct several variants of the Transformer-AE (T-AE). Specifically, T-AE (no env, no Q) excludes both environmental inputs and DQN-based layer selection; T-AE (env, no Q) incorporates environmental features but uses a fixed Transformer depth; T-AE (no env, Q) applies adaptive depth selection without environmental information; and T-AE (env \& Q) combines both mechanisms and corresponds to our AEAT-AE model. These variants allow us to isolate the impacts of environmental awareness and dynamic computation.
	
	\subsubsection{Impact of Environmental Awareness on System Robustness}
	
	To evaluate how environmental parameters influence the proposed AEAT-AE framework, ablation experiments are conducted by selectively removing environmental inputs and comparing the resulting BER performance.
	All evaluated schemes are implemented under a unified IM/DD signaling abstraction and identical channel conditions, in order to enable a controlled comparison of different error-control and end-to-end encoding strategies rather than full physical-layer co-design.
	
	\begin{figure*}[!t]
		\centering
		\includegraphics[width=7in]{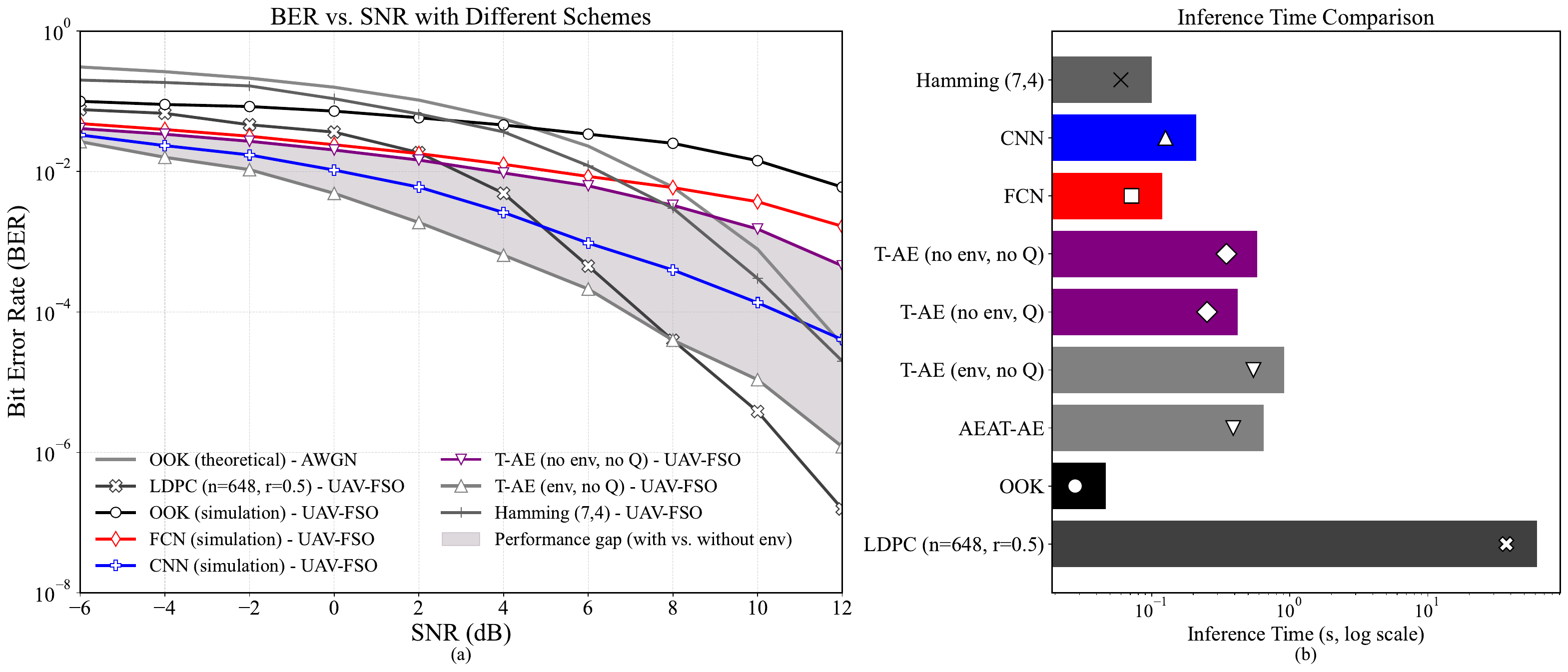}
		\caption{Comparison of different schemes: (a) BER vs. SNR under the same IM/DD channel model, highlighting the impact of environmental awareness; (b) Inference time of these methods for a 200,000-bit stream.}
		\label{BER_vs_SNR}
	\end{figure*}
	
	Fig.~\ref{BER_vs_SNR} summarizes these comparisons. Fig.~\ref{BER_vs_SNR} (a) presents BER versus SNR for the proposed variants, a simple uncoded OOK baseline, neural autoencoders (FCN-AE~\cite{Related1}, CNN-AE~\cite{Related2, Related7}) implemented with fixed architectures following standard practice in prior learning-based physical-layer studies, and classical channel coding schemes including Hamming~(7,4) and LDPC with the standard Wi-Fi block length of 648 bits (rate 1/2). The theoretical BER curve for OOK under AWGN channels is also included for reference. Fig.~\ref{BER_vs_SNR} (b) plots inference latency against BER for a 200,000-bit sequence, with a logarithmic x-axis to reflect the wide dynamic range.

	Among the baselines, uncoded OOK achieves the shortest inference time due to the absence of both encoding and decoding, but its BER deteriorates substantially under UAV–FSO channel impairments. 
	Hamming~(7,4) provides moderate error-correction capability, yielding better BER than uncoded OOK, but its performance is limited compared to LDPC. For classical coded schemes, LDPC provides strong error-correction capability but typically relies on long block lengths to approach its theoretical performance limits, resulting in higher decoding complexity and latency. FCN and CNN autoencoders outperform the uncoded baseline, particularly at low SNRs, with CNN offering improved BER at the cost of moderately increased inference time.
	
	Our Transformer-AE variants achieve an effective trade-off between communication performance and computational cost. Without environmental inputs, the baseline Transformer-AE attains BER performance between that of the FCN- and CNN-based models. When environmental features are incorporated through cross-attention and residual fusion, and all models are trained for the same number of epochs, the BER is reduced by an average of 76.25\% relative to the baseline Transformer-AE, as highlighted by the purple region in Fig.~\ref{BER_vs_SNR} (a). This confirms that environmental awareness provides a substantial robustness gain, further supported by ablation results showing marked degradation when these inputs are removed.
	
	From a theoretical perspective, the computational complexity of the considered architectures scales differently with the sequence length and embedding dimension. For a Transformer input sequence of length $T$ (number of tokens) and embedding dimension $d_k$, the per-layer FLOPs can be summarized as follows:
		\begin{itemize}
			\item Transformer-AE: $O(T^2 d_k + T d_k^2)$, where the quadratic term arises from multi-head self-attention,
			\item CNN-AE: $O(T d_k k)$, scaling linearly with $T$ for a fixed kernel size $k$,
			\item FCN-AE: $O(T d_k^2)$, linear in $T$ but quadratic in the embedding dimension.
		\end{itemize}
	
	In our experiments (sequence length corresponding to 200{,}000 transmitted bits after tokenization and embedding dimension $d_k = 32$), the attention-related quadratic term increases the computational cost of the Transformer-AE compared to CNN and FCN baselines, but does not dominate the overall inference latency. As a result, although the Transformer-AE incurs higher runtime, Fig.~\ref{BER_vs_SNR} (b) shows that the measured inference latencies of the Transformer-AE, CNN, and FCN still remain within the same numerical order of magnitude.

	Furthermore, the proposed AEAT-AE does not reduce computational cost by accelerating individual attention operations, but by adaptively controlling the number of active Transformer layers. Specifically, during inference the effective complexity scales with $O(L_{\text{active}} \cdot (T^2 d_k + T d_k^2))$, where $L_{\text{active}} \leq L$ denotes the number of layers selected by the DQN-based controller. The DQN itself incurs only $O(1)$ overhead with respect to the sequence length, as policy recomputation is triggered infrequently based on slow-varying environmental changes. This design ensures that computational savings arise from reduced layer utilization rather than additional neural processing, yielding a favorable trade-off between BER performance and latency.

	\subsubsection{DQN-Based Adaptive Layer Selection for Efficient Inference}

	\begin{figure*}[!t]
		\centering
		\includegraphics[width=7in]{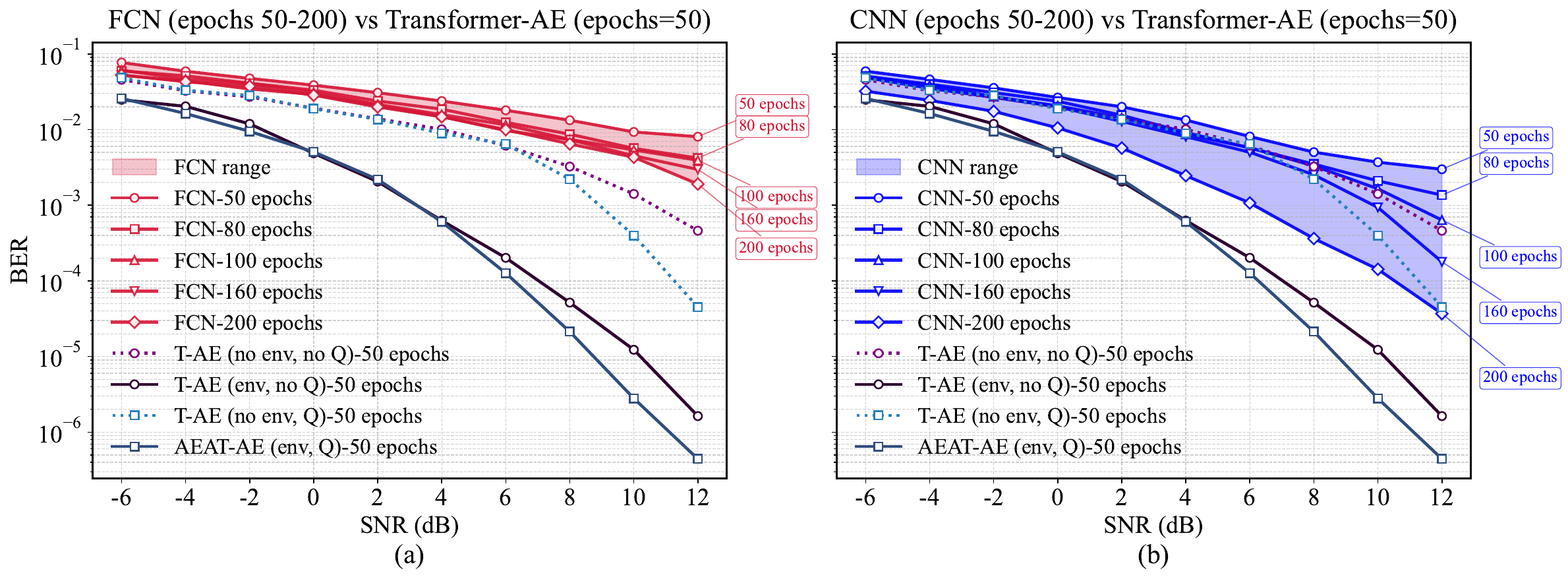}
		\caption{BER vs. SNR at different training epochs for (a) FCN and Transformer-AE, (b) CNN and Transformer-AE.}
		\label{Epoch}
	\end{figure*}

	In our proposed AEAT-AE, the Transformer-AE demonstrates strong feature extraction capabilities and fast convergence. As shown in Fig.~\ref{Epoch}, with the same fixed layer configuration, it achieves comparable or better BER performance than FCN- and CNN-based autoencoders (evaluated over 50 to 200 epochs), despite being trained for only 50 epochs. The base Transformer-AE is first trained for 50 epochs, after which the DQN controller is trained with the autoencoder weights fixed. Curves labeled with ``Q'' (e.g., ``T-AE (env, Q)'') reflect inference-time performance using the trained DQN agent, which dynamically selects a subset of Transformer layers based on environmental inputs. Inactive layers are bypassed via identity mappings, allowing selective activation that modulates computational complexity, which can be quantitatively measured by the average number of active layers.

\begin{figure}[!t]
		\centering
		\includegraphics[width=3.5in]{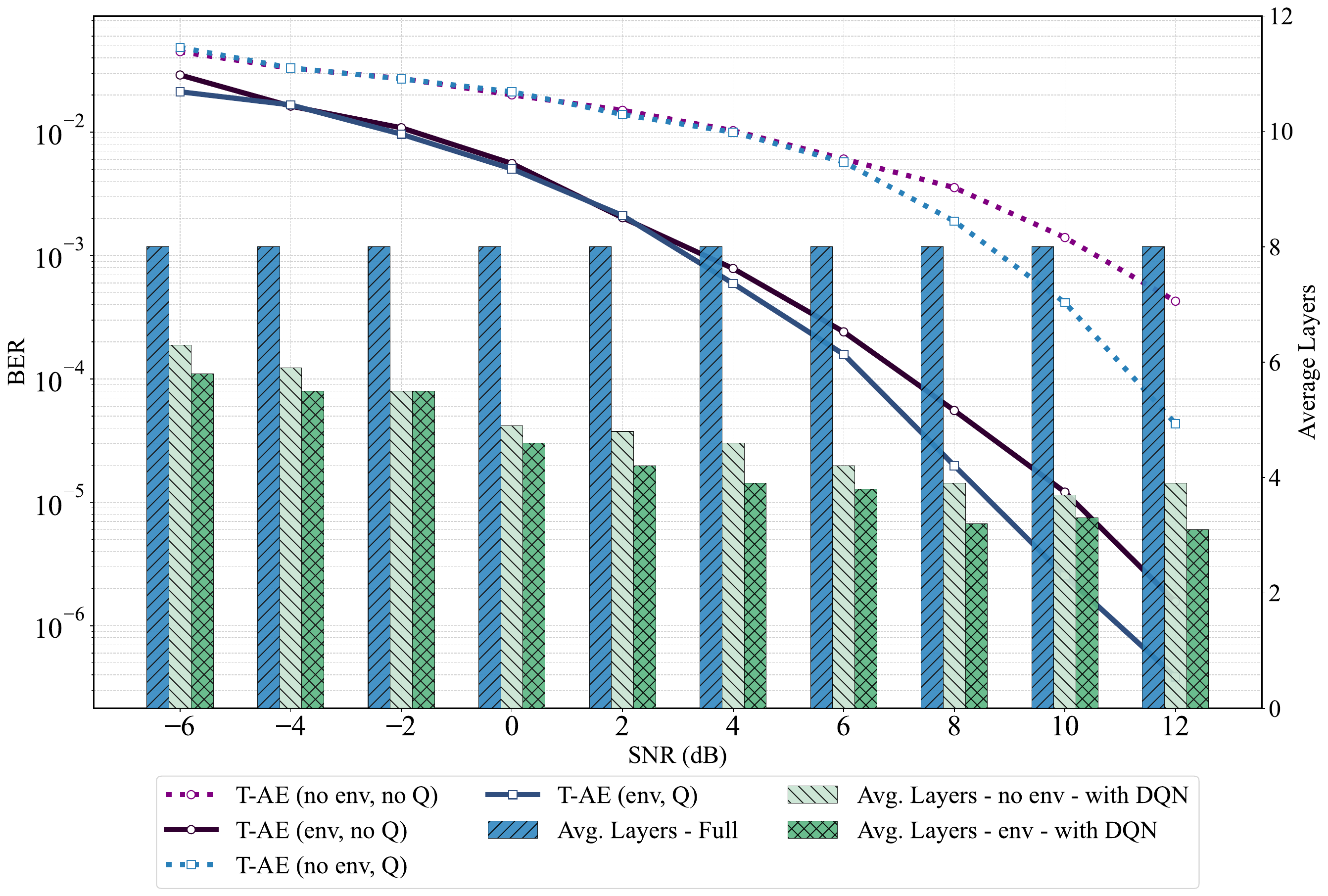}
		\caption{BER performance and average number of active layers under different configurations.}
		\label{Q1}
	\end{figure}

	Fig.~\ref{Q1} compares the BER and the average number of active layers for the four model variants under different SNR conditions. The full-depth Transformer-AE contains 4 encoder and 4 decoder layers (8 layers in total) and serves as the baseline. For the DQN-enabled variants (denoted with ``Q''), the results are averaged over inference episodes after the DQN policy has converged, thereby reflecting its stable adaptive behavior. On average, the DQN agent reduces the number of activated layers by 4.29 without environmental information and by 4.97 when environmental features are included, indicating substantial computational savings.
	
	Moreover, the results further show that the DQN agent activates more layers at low SNR to preserve accuracy and fewer layers at high SNR to reduce computation without degrading BER. Notably, around 6~dB SNR, the adaptive model achieves both improved BER and reduced layer usage compared to the fixed full-depth baseline. This indicates that adaptive layer control effectively balances robustness and efficiency, while implicitly regularizing model capacity by suppressing redundant layers under favorable conditions. This adaptive suppression of unnecessary layers under favorable channel conditions effectively acts as a structural regularizer, improving generalization by preventing overfitting and ensuring that model capacity is aligned with input complexity. From an optimization perspective, such capacity modulation provides an intuitive explanation for the faster convergence observed in Fig.~\ref{Epoch}, as it leads to a smoother optimization landscape and more efficient training than fixed-depth architectures.

	\begin{figure*}[!t]
		\centering
		\includegraphics[width=5.5in]{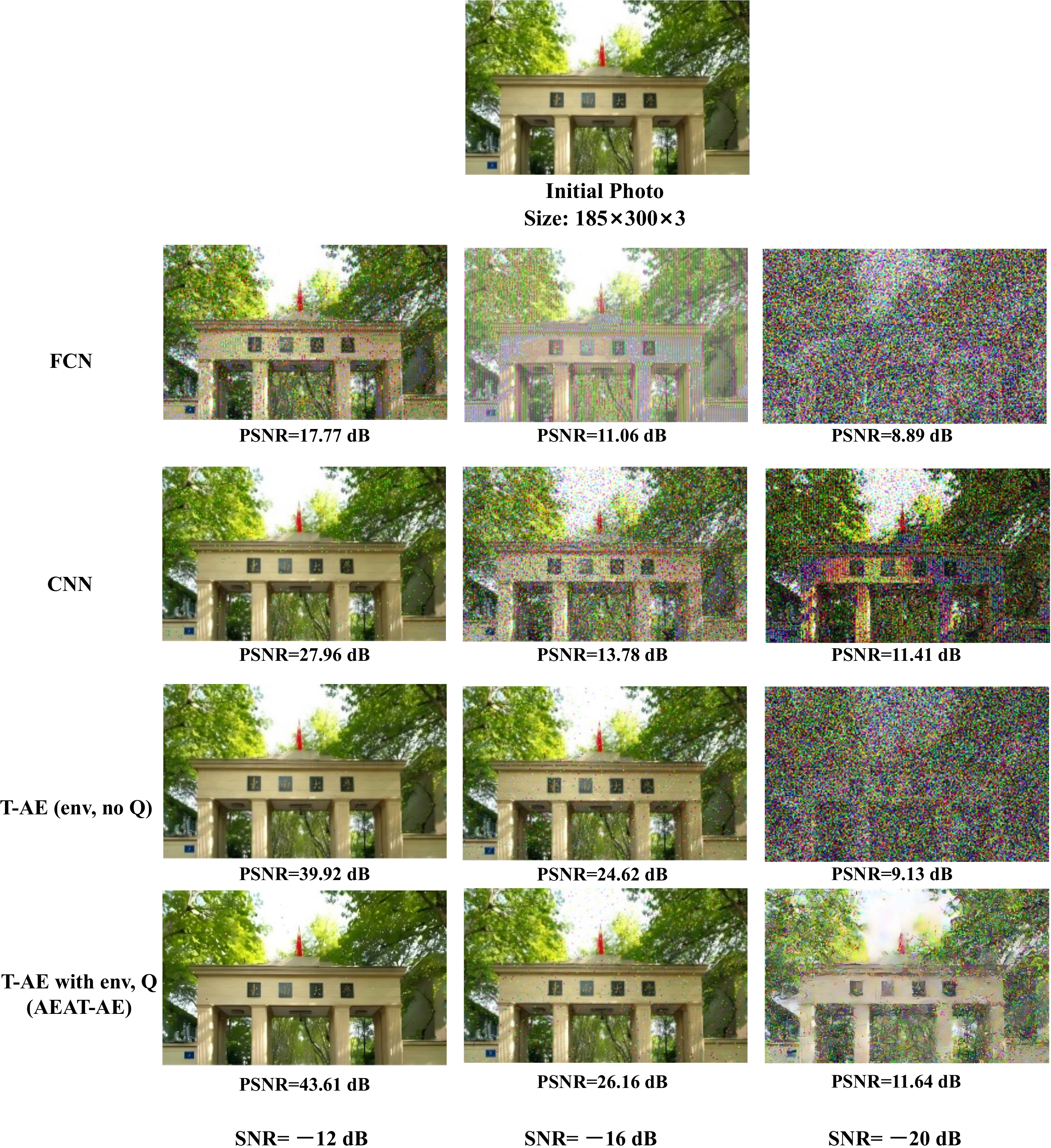}
		\caption{Image transmission performance under different schemes.}
		\label{psnr}
	\end{figure*}
	
	\begin{figure}[!t]
		\centering
		\includegraphics[width=3.2in]{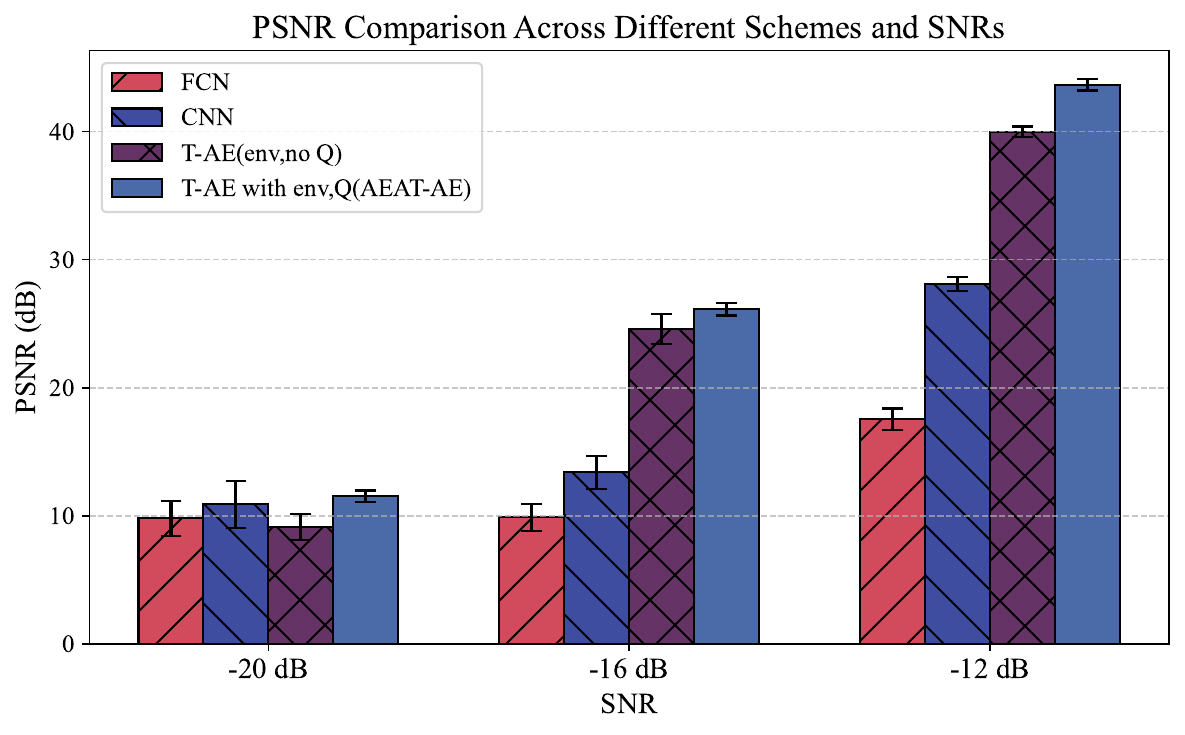}
		\caption{PSNR comparison across different schemes and snrs}
		\label{psnr_bar}
	\end{figure}

	\subsubsection{Image Transmission Performance Under Severe Channel Conditions}
	
	To intuitively demonstrate the practical effectiveness of the proposed method, as shown in Fig. \ref{psnr}, we present a visual case study of image transmission and reconstruction over a UAV-FSO channel under adverse conditions. For consistency across experiments, the input image is converted to three-channel RGB, composited onto a white background when transparency is present, quantized to 8 bits per channel, and serialized into a binary bitstream. The bitstream is zero-padded to a fixed length and segmented into equal-sized blocks before being modulated and transmitted through the transmitter-side autoencoder over the simulated UAV–FSO channel.
	
	At the receiver, the decoder output is thresholded at 0.5 to regenerate the binary bitstream, which is then truncated to the original length prior to padding. The bitstream is decoded back into 8-bit pixel values and reshaped to reconstruct the original image dimensions. Reconstruction quality is evaluated by the Peak Signal-to-Noise Ratio (PSNR), defined as:
	
	\begin{equation}
		\operatorname{PSNR} = 10 \cdot \log_{10}\left(\frac{\mathrm{MAX}^2}{\mathrm{MSE}}\right),
	\end{equation}
	
	\noindent where \(\mathrm{MAX} = 255\) is the maximum pixel value, and \(\mathrm{MSE}\) is the mean squared error between the original and reconstructed images. Higher PSNR indicates better visual fidelity.
	
	As shown in Fig.~\ref{psnr_bar}, the average PSNR and its variance are computed from multiple independent transmission trials (noise realizations) of the same input image under the same channel conditions, and are reported for different schemes under varying SNR conditions. Under extremely low SNR conditions, the FCN-based scheme yields the lowest average PSNR with large performance fluctuations, indicating severe degradation in reconstruction quality. The CNN-based model provides modest improvements in terms of mean PSNR, but still suffers from noticeable performance loss, particularly at SNR values of $-16$~dB and $-20$~dB. In comparison, the environment-aware Transformer-based approaches—T-AE (env, no Q) and the proposed AEAT-AE—achieve consistently higher mean PSNR with reduced variance, demonstrating stronger robustness to channel impairments. Both variants benefit from environmental inputs, while AEAT-AE further incorporates DQN-driven adaptive layer selection. This allows it to match or surpass the performance of T-AE (env, no Q) while activating fewer layers, leading to a more favorable trade-off between reconstruction quality and computational efficiency. These results suggest that AEAT-AE provides a balanced and stable performance under challenging SNR conditions.

	\section{Discussion}
	
	This study uses comprehensive, controlled simulations as a necessary first step in UAV-FSO research, serving as a pilot investigation that demonstrates the feasibility and effectiveness of the AEAT-AE framework with DQN-based adaptive layer selection. Our focus is on validating design and adaptability under diverse channel conditions, laying groundwork for future real-world deployment and refinement.
	
	To achieve such controlled validation, several simplifying assumptions were made in the simulation setup. In particular, the atmospheric turbulence along the UAV-FSO link is considered homogeneous. While this assumption facilitates repeatable and interpretable evaluation of the proposed framework, it does not fully capture the non-uniform turbulence that typically arises from altitude and spatial variations in practical deployments. This idealization is acceptable for model verification but inevitably limits the realism of the results. Future work will extend the framework to heterogeneous atmospheric conditions and incorporate empirical measurements to assess robustness under more complex environments.
	
	Beyond the channel modeling assumptions, given practical systems' stringent real-time and complexity constraints, further work on more lightweight adaptive strategies is warranted. While deep reinforcement learning offers enhanced adaptability, the current DQN strikes a practical balance between performance and computational cost suitable for edge deployment. Future research may consider more advanced DRL methods if they offer clear gains in adaptability or control precision.

	\section{Conclusion}
	
	In this work, we introduced the Adaptive Environment-aware Transformer-AE (AEAT-AE) framework for UAV-FSO communication, which leverages environmental parameters and reinforcement learning to enhance both robustness and efficiency. By incorporating environmental awareness through cross-attention and introducing a DQN-driven adaptive layer selection mechanism, AEAT-AE can adapt to varying channel conditions while reducing computational complexity. Extensive simulations demonstrate notable improvements in key performance metrics, including BER, PSNR, and layer efficiency. The results suggest the practical feasibility of AEAT-AE for UAV-FSO deployments, highlighting its potential for next-generation adaptive optical wireless communication systems operating efficiently under dynamic and fluctuating conditions.

	\vfill
	
\end{document}